\documentclass[amsmath,amssymb,aps,prd,preprintnumbers,superscriptaddress,eqsecnum,showpacs,nofootinbib,showkeys
,twocolumn 
]{revtex4-2} 
\usepackage{blkarray,graphicx,dcolumn,bm,hyperref,physics,here,mathrsfs,cases,empheq,latexsym,amsfonts,color,listings,subcaption,orcidlink}

\begin{document}

\allowdisplaybreaks[1]
\title{Resonance in black hole ringdown:\\ Benchmarking quasinormal mode excitation and extraction}

\author{Kei-ichiro Kubota\orcidlink{0000-0002-1576-4332}}
    \affiliation{Institute for Cosmic Ray Research, The University of Tokyo, 5-1-5 Kashiwanoha, Kashiwa, Chiba 277-8582, Japan}
\author{Hayato Motohashi\orcidlink{0000-0002-4330-7024}}
    \affiliation{Department of Physics, Tokyo Metropolitan University, 1-1 Minami-Osawa, Hachioji, Tokyo 192-0397, Japan}
\date{\today}
\begin{abstract}
We investigate how resonant excitation near exceptional points manifests in Kerr black hole ringdown waveforms and examine its extraction.
Using waveforms generated by localized initial data, where quasinormal mode amplitudes are given solely by excitation factors, we establish a controlled benchmark for overtone extraction. 
Applying an iterative fitting method with mirror modes, we analyze a mild resonance in the $(l,m)=(2,2)$ multipole and a sharp resonance in the $(3,1)$ multipole occurring as part of a sequence of successive resonances. 
For $(2,2)$, we extract the fundamental mode, the first three overtones, and the fundamental mirror mode with relative errors below $10\%$, and show that residual waveforms exhibit the expected damped sinusoids together with distinctive resonance signatures. 
For $(3,1)$, we demonstrate that resonances can not only \emph{amplify} but also \emph{reduce} quasinormal mode excitations, reshaping the overtone hierarchy and rendering the sharp resonance more pronounced in ringdown. 
Our results clarify the imprint of resonance in Kerr ringdown and highlight both the robustness and limitations of current extraction techniques, providing a foundation for more reliable extraction of higher overtones and for applications to observational data analysis.
\end{abstract}
\maketitle

\section{\label{sec:intro}Introduction}

The ringdown phase of gravitational waves emitted by perturbed black holes encodes fundamental information about the spacetime geometry in the strong-field regime~\cite{LIGOScientific:2025obp,Maenaut:2024oci,Chung:2025wbg} (see reviews~\cite{Berti:2009kk,Hatsuda:2021gtn,Berti:2025hly}).
In this phase, the signal is well described by a superposition of quasinormal modes (QNMs), whose complex frequencies are determined solely by the background spacetime.
In addition to the QNM frequencies, their complex amplitudes also carry key information about the strong-field physics, such as the excitation process and the spacetime structure, because the amplitudes can be factorized into a term dependent on the initial data and an excitation factor, the latter of which depends only on the background spacetime.

Recent studies have revealed that the excitation factors play a dominant role in determining the overtone hierarchy, particularly for prograde modes in ringdown signals.
This behavior has been confirmed through fitting analyses of the ringdown signal in multiple contexts: waveforms sourced by plunging particles~\cite{Berti:2006wq,Oshita:2021iyn}, numerical relativity simulations of binary black hole mergers~\cite{Cheung:2023vki,Giesler:2024hcr,Mitman:2025hgy} (see also Sec.~5.1.3 in Ref.~\cite{Berti:2025hly}), and direct evaluations of the source term in the Schwarzschild case~\cite{Zhang:2013ksa}.~\footnote{However, the ratio between the prograde mode and retrograde mode depends on the initial data~\cite{Cheung:2023vki}.}
While the absolute amplitude depends on the initial data, the relative ratios among prograde overtones are largely governed by the excitation factors and exhibit only a weak dependence on the initial data and source configuration.
Consequently, a reliable extraction of QNM amplitudes across multiple overtones would open a new window for probing the strong-gravity field and diagnosing the black hole parameters beyond what is accessible from frequency measurements alone.

A particularly intriguing feature arises when QNM frequencies approach each other in the complex plane.
This leads to an avoided crossing in which the QNM frequencies trace a hyperbolic trajectory, together with resonant amplification of the corresponding excitation factors in a lemniscate pattern~\cite{Motohashi:2024fwt,motohashi_2024_12696858}, which has been proposed recently as a resolution of the longstanding mystery of the ``dissonance'' anomaly in Kerr QNM spectrum~\cite{Onozawa:1996ux}, as well as a novel diagnostic of black hole structure and a potential probe of new physics.
Furthermore, it has been derived in Ref.~\cite{Motohashi:2024fwt} that this phenomenon universally occurs near exceptional points~\cite{Kato1995}, a characteristic feature of non-Hermitian systems that were identified in gravity for the first time.
Similar behavior has also been reported in Einstein-Maxwell-axion system~\cite{Takahashi:2025uwo}, reinforcing the potential of the resonance to probe new physics.

From the symmetric amplification of excitation factors, one might expect that such paired modes with destructive interference do not cause obvious signatures and only mildly affect the ringdown signal~\cite{Oshita:2025ibu}. 
However, in general, exceptional points generate a qualitatively different time dependence in non-Hermitian systems~\cite{Heiss2010,Ashida:2020dkc}.
Such a peculiar time dependence has been indeed confirmed in the ringdown waveform through fitting, as demonstrated for the resonance involving the fundamental mode in the perturbed Regge-Wheeler potential~\cite{Yang:2025dbn}.
Still, since resonance phenomena in Kerr QNMs occur between higher overtones, it remains unclear both how such resonances manifest in the ringdown waveform and how reliably they can be extracted.

In general, the extraction of overtone amplitudes remains a challenging problem.  
Traditional fitting approaches face ambiguities in determining the ringdown start time~\cite{Nollert:1996rf,Thrane:2017lqn,Bhagwat:2017tkm} and are prone to overfitting~\cite{Baibhav:2023clw,Nee:2023osy,Ma:2022wpv,Ma:2023cwe,Takahashi:2023tkb}.
The standard mismatch metric is often insufficient to guarantee that the fit captures the true physical content of the signal~\cite{Baibhav:2023clw,Nee:2023osy}.
The practical way to evaluate the quality of the fit is to examine the stability of the parameters when varying the fitting window.
While it depends on the fitting method, criteria of stable extraction, and the waveform of interest, the robust fitting of overtones has so far been limited to roughly the second~\cite{Zhu:2023mzv,Gao:2025zvl}, third~\cite{Clarke:2024lwi,Mitman:2025hgy}, or fourth overtone~\cite{Takahashi:2023tkb}.
Developing methods to reliably fit higher overtones, as well as identifying better indicators of fit quality, remains an important and pressing task.

One promising approach is the iterative fitting algorithm proposed in Ref.~\cite{Takahashi:2023tkb}, which extracts QNM contributions sequentially and mitigates the overfitting issue.
In this algorithm, we first identify the longest-lived QNM in the waveform and subtract it from the data.
The resulting residual is dominated by the next-longest-lived mode.
By iteratively repeating this process, we can systematically peel off one mode after another.
The iterative method has shown promise in extracting multiple overtones with improved stability~\cite{Takahashi:2023tkb}.
This strategy has been employed in both linear and nonlinear fitting in recent studies~\cite{Clarke:2024lwi,Giesler:2024hcr,Gao:2025zvl,Mitman:2025hgy}. 

The aim of the present work is to identify and characterize the features of resonant excitation in the Kerr ringdown waveform. 
Our analysis combines three key ingredients: recently computed excitation factors~\cite{Motohashi:2024fwt,motohashi_2024_12696858}, the iterative extraction method~\cite{Takahashi:2023tkb}, and a Kerr ringdown waveform generated by localized initial data~\cite{Andersson:1996cm,Berti:2006wq}. 
In this waveform, modeled as a delta-function source term, the QNM amplitudes are given solely by the excitation factors.
This setup provides a controlled and idealized environment for exploring resonance features and testing their extraction from a waveform composed of an infinite QNM sum and power-law tails. 
While the response of a localized source has been studied in previous work~\cite{Andersson:1996cm,Berti:2006wq,Oshita:2024wgt,Oshita:2025ibu}, to our knowledge this is the first time that the fitting of the resulting waveform is explicitly performed.
Since QNM amplitudes in more realistic waveforms are also largely governed by excitation factors, the insights gained from this analysis can be directly applied to the fitting of observational signals.

The rest of this paper is organized as follows.  
In Sec.~\ref{sec:ringdown_waveform}, we summarize the Teukolsky formalism, the QNM decomposition of the ringdown, and the waveform we investigate.
We also address the ambiguities in the definition of excitation factors, which can sometimes be a source of confusion in practice.
In Sec.~\ref{sec:method}, we describe the iterative fitting algorithm of QNMs as well as the fitting of the power-law tail.
We present the results of our analysis of mild and sharp resonances found in Kerr QNMs recently in Sec.~\ref{sec:results}, including a detailed analysis of extraction stability, the comparison with the excitation factors, and characteristic resonance features.
Finally, we conclude with a summary and outlook in Sec.~\ref{sec:conclusion}.  
Throughout this work, we adopt geometrical units $c = G = 1$ and use Boyer-Lindquist coordinates.

\section{\label{sec:ringdown_waveform}Ringdown and QNMs}

In this section, we first review the Teukolsky formalism for linear perturbations of Kerr black holes, and the QNM decomposition of the ringdown. 
We also clarify the ambiguities in the definition of excitation factors.
We then introduce the waveform induced by a localized initial data, which we mainly investigate in the present paper.

\subsection{\label{sec:Teukolsky}Teukolsky formalism}
Gravitational waves emitted in the ringdown phase of binary black hole mergers are described by linear perturbations on a background black hole spacetime (see \cite{Nakamura:1987zz,Sasaki:2003xr} for reviews). The perturbations on the Kerr spacetime are governed by a master equation called the Teukolsky equation~\cite{Teukolsky:1973ha,Press:1973zz,Teukolsky:1974yv} derived within the Newman-Penrose formalism~\cite{Newman:1961qr}. The master variable for the Teukolsky equation is $\rho^{-4}\Psi_4$, where $\rho=(r-ia\cos\theta)^{-1}$ and $\Psi_4$ is a component of the Weyl tensor projected onto null tetrads in the Newman-Penrose formalism. In the asymptotic region far from the perturbed Kerr black hole, $\Psi_4$ is related to the gravitational wave strain by $\Psi_4 = \frac{1}{2} \frac{\mathrm{d}^2}{\mathrm{d}t^2}(h_+ -i h_\times) $ at infinity~\cite{Nakamura:1987zz}. Decomposing the master variable by the Laplace transform
\begin{align}
    \rho^{-4}\Psi_4 = \sum_{l=2}^{\infty}\sum_{m=-l}^l\frac{1}{\sqrt{2\pi}}\int\mathrm{d}\omega e^{-i\omega t + im\varphi} R_{lm}(r)S_{lm}(\theta;a\omega),
    \label{eq:decomposition}
\end{align}
the Teukolsky equation reduces to separate radial and angular equations:
\begin{align}
    &\Biggl[\Delta^{2}\frac{\mathrm{d}}{\mathrm{d}r}\biggl(\Delta^{-1}\frac{\mathrm{d}}{\mathrm{d}r}\biggr) + \frac{K^2+4i(r-M)K}{\Delta}\nonumber \\
    &-8i\omega r-\lambda\Biggr]R_{lm}(r) = T_{lm}(r),\\
    &\Biggl[\frac{1}{\sin\theta}\frac{\mathrm{d}}{\mathrm{d}\theta}\left(\sin\theta\frac{\mathrm{d}}{\mathrm{d}\theta}\right)-a^2\omega^2\sin^2\theta -\frac{(m-2\cos\theta)^2}{\sin^2\theta}\nonumber \\
    &+4a\omega \cos\theta-2+2a\omega m+\lambda\Biggr]S_{lm}(\theta;a\omega)=0,
\end{align}
where $K=(r^2+a^2)\omega - ma $ and $\Delta=r^2-2Mr+a^2=(r-r_+)(r-r_-)$ with $r_\pm=M\pm\sqrt{M^2-a^2}$.
$\lambda$ is the separation constant, and $T_{lm}$ is the source term (see \cite{Sasaki:2003xr} for detail expression). The angular function is normalized as
\begin{align}
    \int_0^\pi S^*_{lm}(\theta;c) S_{lm}(\theta;c)\sin\theta\mathrm{d}\theta = 1,
\end{align}
with $c=a\omega$.

\begin{align}
    R^\text{in}_{lm}\to
    \left\{ \,
        \begin{aligned}
            &\Delta^2e^{-ikr_*} & \text{for } r\to r_+\\
            &r^3B^\text{ref}_{lm}e^{i\omega r_*} + r^{-1}B^\text{inc}_{lm}e^{-i\omega r_*} & \text{for } r\to \infty,
        \end{aligned}
    \right.\label{eq:Rinasymp}
\end{align}
and the ``up'' solution $R^\text{up}_{lm}$ satisfying 
\begin{align}
    R^\text{up}_{lm}\to
    \left\{ \,
        \begin{aligned}
            &C^\text{up}_{lm}e^{ikr_*} + \Delta^2C^\text{ref}_{lm}e^{-ik r_*}  & \text{for } r\to r_+\\
            &r^3e^{i\omega r_*} & \text{for } r\to \infty,
        \end{aligned}
    \right.\label{eq:Rupasymp}
\end{align}
where $B^\text{inc}_{lm}$, $B^\text{ref}_{lm}$, $C^\text{inc}_{lm}$, and $C^\text{ref}_{lm}$ are complex asymptotic amplitudes and $k = \omega - am/(2Mr_+)$. $r_*$ is the tortoise coordinate which satisfies $\mathrm{d}r_*/\mathrm{d}r=(r^2+a^2)/\Delta$.
We fix the integration constant and employ the conventional tortoise coordinate,
\begin{align}\label{eq:tortoise}
    r_* = r + \frac{2Mr_+}{r_+ - r_-}\ln\frac{r-r_+}{2M} - \frac{2Mr_-}{r_+ - r_-}\ln\frac{r-r_-}{2M}.
\end{align}
Other choices generate a complex exponential factor in the asymptotic amplitudes (see Sec.~\ref{sec:remark}).

The radial Green's function is then given by 
\begin{align}
    G_{lm}(r,r') =& \frac{\Delta^{-2}(r')}{W[R^\text{in}_{lm},R^\text{up}_{lm}]} \biggl[R^\text{in}_{lm}(r')R^\text{up}_{lm}(r)\theta(r-r')\nonumber \\
    & + R^\text{in}_{lm}(r)R^\text{up}_{lm}(r')\theta(r'-r)\biggr],
    \label{eq:Greenfunction}
\end{align}
where $\theta(x)$ is the Heaviside step function and $W[R^\text{in}_{lm},R^\text{up}_{lm}]$ is the Wronskian defined by 
\begin{align}
    W[R^\text{in}_{lm},R^\text{up}_{lm}] &= \Delta^{-1}\biggl(R^\text{in}_{lm} \frac{\mathrm{d}R^\text{up}_{lm}}{\mathrm{d}r} - R^\text{up}_{lm} \frac{\mathrm{d}R^\text{in}_{lm}}{\mathrm{d}r}\biggr).
\end{align}
Substituting the asymptotic solution~\eqref{eq:Rinasymp} and \eqref{eq:Rupasymp} at infinity into this expression, we obtain
\begin{align}
    W[R^\text{in}_{lm},R^\text{up}_{lm}] = 2i\omega B^\text{inc}_{lm}.
\end{align}
The Wronskian vanishes when $R^\text{in}_{lm}$ and $R^\text{up}_{lm}$ become linearly dependent. 
In this case, the solution simultaneously satisfies the ingoing boundary condition at the horizon ($r\to r_+$) and the outgoing boundary condition at infinity ($r\to \infty$).
This occurs only for a discrete set of complex frequencies, which define the QNM spectrum.
Equivalently, the QNM frequencies are determined by the condition $B^\text{inc}_{lm}(\omega) = 0$, and are identified as the poles of the Green's function.

Using the Green's function~\eqref{eq:Greenfunction}, the inhomogeneous solution is given by 
\begin{align}
    R_{lm}(r) =& \int_{r_+}^{\infty}\mathrm{d}r' G_{lm}(r,r')T_{lm}(r')\nonumber \\
    =& \frac{1}{2i\omega B^\text{inc}_{lm}}\Biggl[R^\text{up}_{lm}(r)\int_{r_+}^{r}\mathrm{d}r'\Delta^{-2}R^\text{in}_{lm}T_{lm}\nonumber \\
    &+ R^\text{in}_{lm}(r)\int_{r}^{\infty}\mathrm{d}r'\Delta^{-2}R^\text{up}_{lm}T_{lm}\Biggr].
\end{align}
Therefore, the asymptotic behavior at infinity is
\begin{align}
    R_{lm}(r\to\infty) = \frac{r^3e^{i\omega r_*}}{2i\omega B^\text{inc}_{lm}} \int_{r_+}^{\infty}\mathrm{d}r'\Delta^{-2}R^\text{in}_{lm}T_{lm}.
\end{align}
Substituting this into Eq.~\eqref{eq:decomposition}, we obtain
\begin{widetext}
\begin{align}
    \Psi_4(t,r\to\infty,\theta,\phi) = - \frac{1}{2\pi}\frac{1}{r}\sum_{l=2}^{\infty}\sum_{m=-l}^l\int^\infty_{-\infty}\mathrm{d}\omega  \frac{B^\text{ref}_{lm}}{2i\omega B^\text{inc}_{lm}} I_{lm} S_{lm}(\theta;a\omega)\frac{e^{im\varphi}}{\sqrt{2\pi}} e^{-i\omega (t-r_*)}
    \label{eq:Psi4atinfinity}
\end{align}
\end{widetext}
where $I_{lmn}$ is source-dependent term defined by
\begin{align}
    I_{lm} =  - 2\pi\int_{r_+}^{\infty}\mathrm{d}r'\frac{R^\text{in}_{lm}(r')T_{lm}(r')}{B^\text{ref}_{lm}\Delta^{2}(r')}.
    \label{eq:sourcedependnetterm}
\end{align}
Equation~\eqref{eq:Psi4atinfinity} highlights that the behavior of the Green's function on the real frequency axis plays a crucial role in understanding ringdown~\cite{Kyutoku:2022gbr}.

\subsection{\label{}QNM decomposition}
\begin{figure}
    \includegraphics[width=1.0\linewidth]{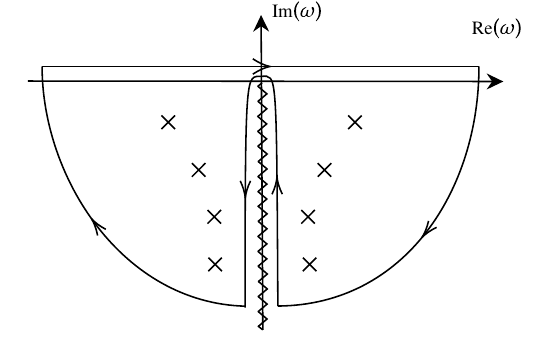}
    \caption{\label{fig:integration_contour} 
    Integration contour for Eq.~\eqref{eq:Psi4atinfinity}. 
    The cross markers and the jagged line denote the QNM poles and the branch cut, respectively. 
    The integral along the real frequency axis in Eq.~\eqref{eq:Psi4atinfinity} is equivalent to closing the contour in the lower half-plane, provided that the contributions from the branch cut and the semicircular arc at infinity are properly taken into account. 
    By the residue theorem, the integral over the closed contour yields the sum over QNMs, while the integral along the branch cut gives rise to the power-law tail. 
    The contribution from the semicircular arc vanishes by Jordan's lemma.}
\end{figure}

While Eq.~\eqref{eq:Psi4atinfinity} involves an integration along the real frequency axis, by enclosing the integration contour on the complex frequency plane, we can rewrite it as a superposition of QNMs together with a power-law tail, provided that no pole arises in the source-dependent term (see Fig.~\ref{fig:integration_contour}).
By evaluating the complex frequency integral using the residue theorem, we obtain
\begin{align}
    \Psi_4 = & \frac{1}{r}\sum_{lmn} I_{lmn}B_{lmn}S_{lm}(\theta;a\omega_{lmn})\frac{e^{im\varphi}}{\sqrt{2\pi}}e^{-i\omega_{lm}(t-r_*)} \nonumber \\ & + \text{(power-law tail)},
    \label{eq:ringdownwaveform}
\end{align}
where $B_{lmn}$ is the excitation factor~\cite{Leaver:1986gd,Andersson:1995zk,Berti:2006wq} defined by
\begin{align}
    B_{lmn} = & \left. \frac{B^\text{ref}_{lm}}{2\omega}\left(\frac{\mathrm{d}B^\text{inc}_{lm}}{\mathrm{d}\omega}\right)^{-1} \right|_{\omega=\omega_{lmn}},
    \label{eq:excitationfactor}
\end{align}
$I_{lmn}$ is the source-dependent term evaluated at the QNM frequency,
\begin{align}
    I_{lmn} = & \left. I_{lm} \right|_{\omega=\omega_{lmn}},
    \label{eq:sourcedependnettermatQNM}
\end{align}
and the index $n$ labels the QNM overtones.

A caveat here is that, while it is generally understood that the ringdown waveform is described by QNMs and the power-law tail, it is worth noting that a distinct damped sinusoidal component—referred to as a ``horizon mode'' or ``redshift mode''—has also been reported~\cite{Mino:2008at,Zimmerman:2011dx,DeAmicis:2025xuh}. 
These modes arise from the pole of the source-dependent term $I_{lm}$, and exhibit a frequency that differs from the conventional QNMs. 
In this work, we shall focus on the ringdown waveform induced by a localized initial data (see Sec.~\ref{sec:waveform}), which does not include the redshift mode.

\begin{figure*}[]
    \begin{minipage}[]{\columnwidth}
        \href{https://zenodo.org/records/18676426/preview/qnm_s-2l2m2.mp4}        {\includegraphics[width=\linewidth]{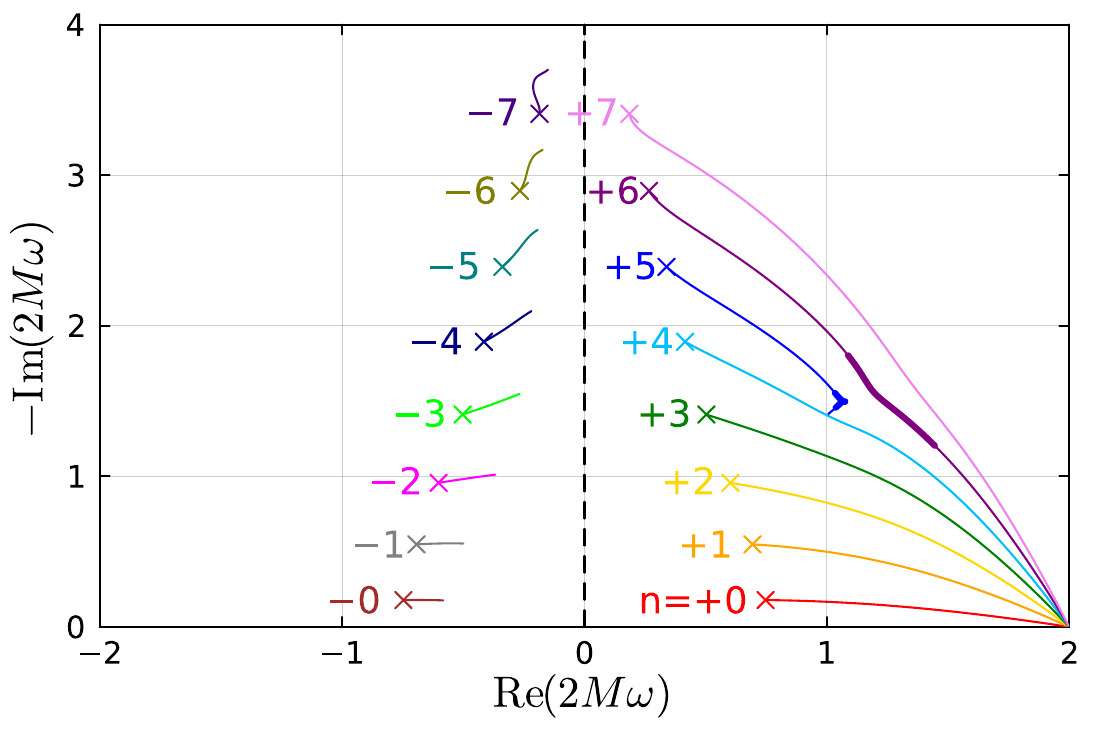}}
        \subcaption{\label{fig:qnm}QNM frequencies for $(l,m)=(2,2)$}
    \end{minipage}
    \begin{minipage}[]{\columnwidth}
        \href{https://zenodo.org/records/18676426/preview/excitation_factor_s-2l2m2_n56.mp4}
        {\includegraphics[width=\linewidth]{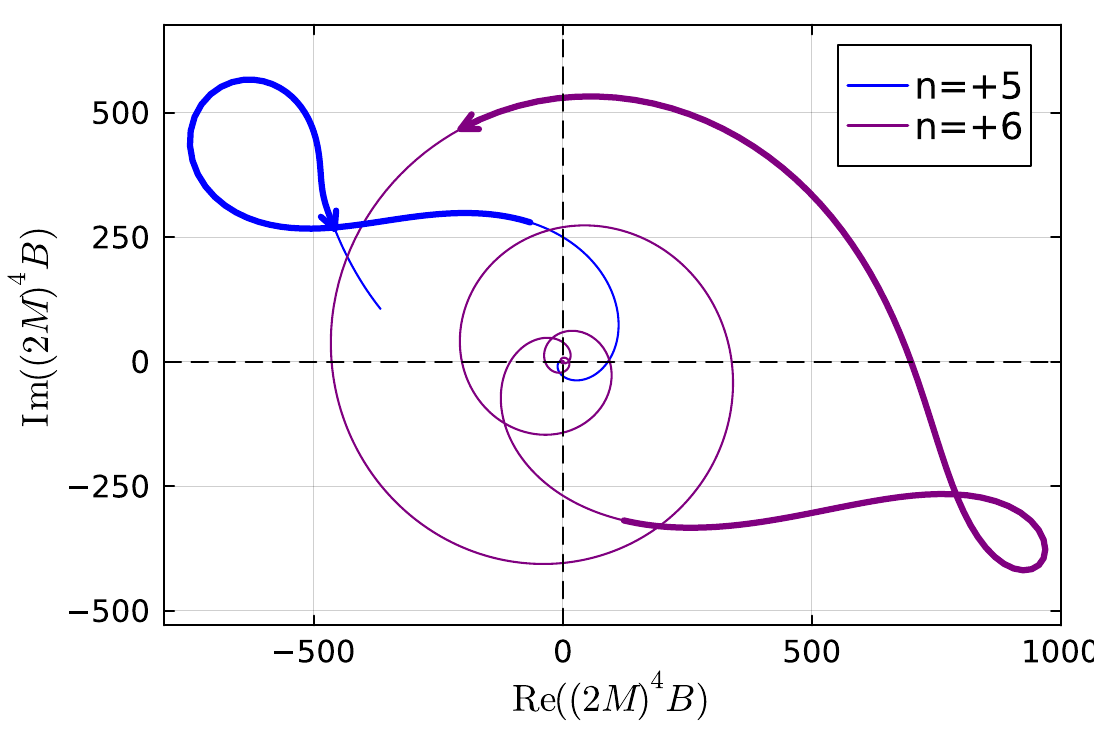}}
        \subcaption{\label{fig:excitationfactor}Excitation factor for $(l,m)=(2,2)$}
    \end{minipage}\\
        \begin{minipage}[]{\columnwidth}
        \href{https://zenodo.org/records/18676426/preview/qnm_s-2l3m1.mp4}
        {\includegraphics[width=\linewidth]{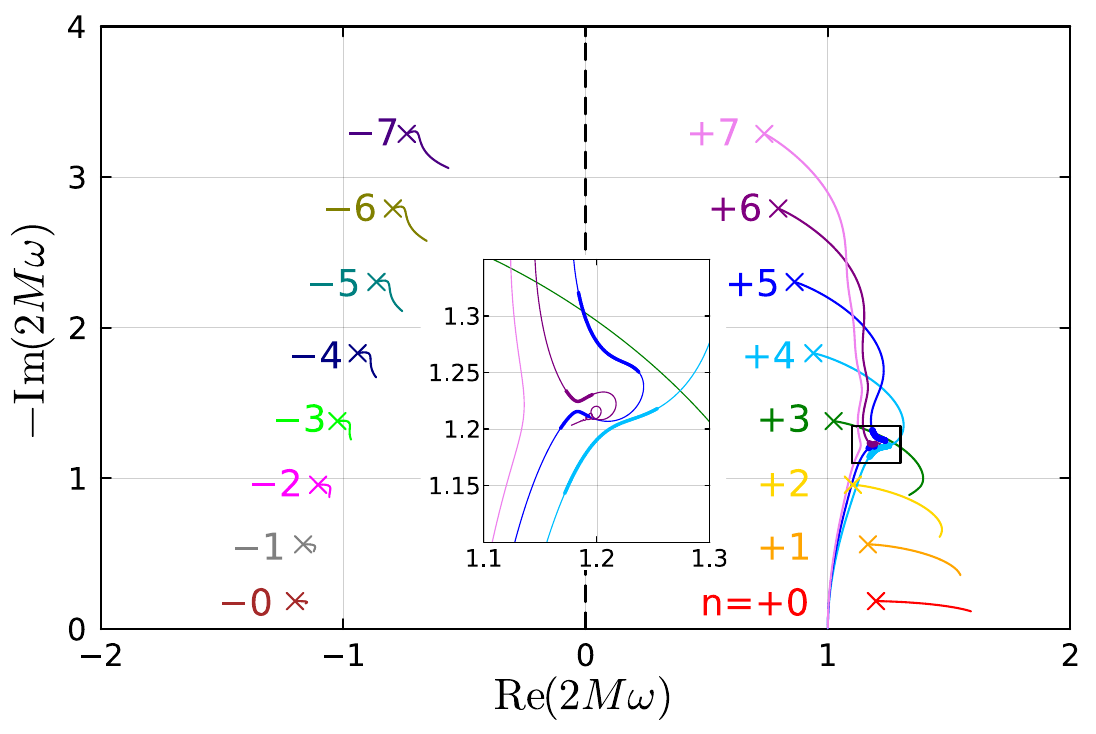}}
        \subcaption{\label{fig:qnm31}QNM frequencies for $(l,m)=(3,1)$}
    \end{minipage}
    \begin{minipage}[]{\columnwidth}
        \href{https://zenodo.org/records/18676426/preview/excitation_factor_s-2l3m1.mp4}
        {\includegraphics[width=\linewidth]{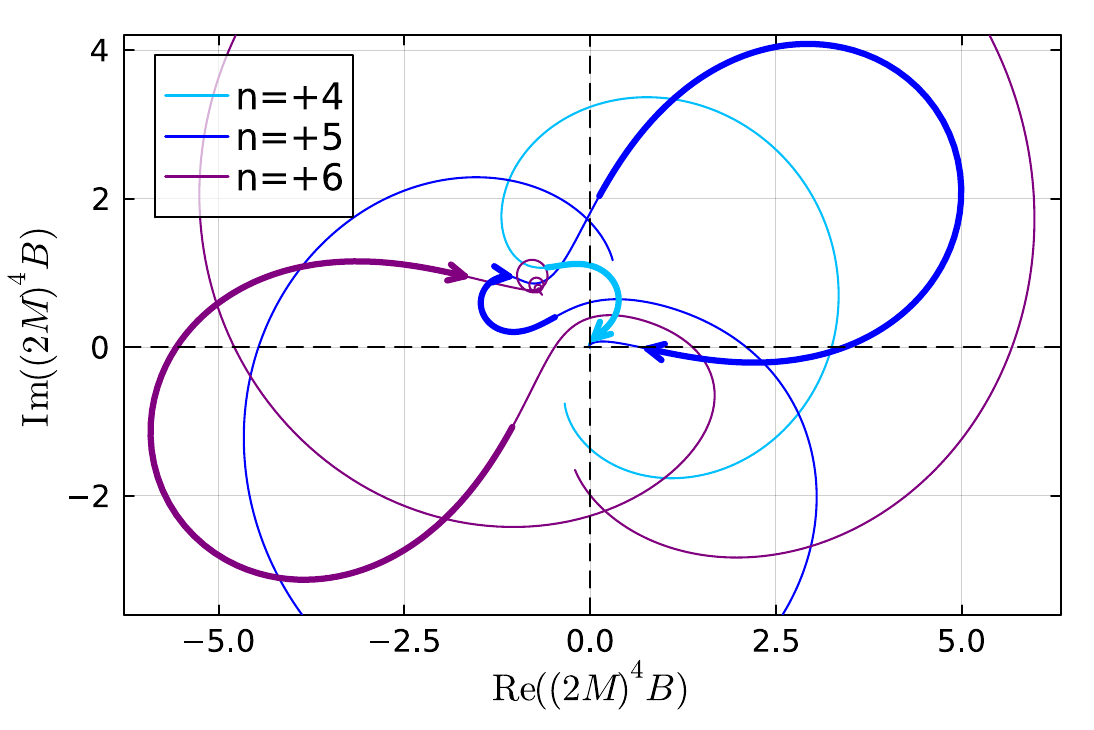}}
        \subcaption{\label{fig:excitationfactor31}Excitation factor for $(l,m)=(3,1)$}
    \end{minipage}
    \caption{\label{fig:qnmfreqandexcitationfactor}
    QNM frequencies (left column) and excitation factors (right column) for the $(l,m)=(2,2)$ mode (top row) and $(l,m)=(3,1)$ mode (bottom row) with spin parameter $0\leq a/M\leq 1 - 10^{-6}$.
    The cross marker in the left column panels denotes the Schwarzschild value. For $(l,m)=(2,2)$, the range $0.85 \leq a/M \leq 0.95$, which corresponds to the repulsion between $n=+5$ and $+6$, is highlighted with thick curves, and the arrow in the right panel shows the direction of increasing spin. For $(l,m)=(3,1)$, the ranges $0.952 \leq a/M \leq 0.96$ and $0.972 \leq a/M \leq 0.99725$, which correspond to the repulsion between $n=+4$ and $+5$ and between $n=+5$ and $+6$ respectively, are highlighted. All figures involving QNM frequencies or excitation factors use the reference dataset (RD)~\cite{Motohashi:2024fwt,motohashi_2024_12696858}.
    Animations of these figures are available in Ref.~\cite{kubota_2026_18676426}.}
\end{figure*}

In what follows, we clarify the notation and conventions regarding positive- and negative-frequency QNMs, characterized by $\Re(\omega)>0$ and $\Re(\omega)<0$, respectively (see Fig.~\ref{fig:qnmfreqandexcitationfactor}). The positive frequency modes are often referred to as ``ordinary modes,'' whereas the negative frequency modes are called ``mirror modes.'' This terminology originates from the fact that, in nonrotating black hole spacetime, the QNM spectrum is symmetric with respect to the imaginary axis, such that the mirror modes are reflections of the ordinary modes, analogous to virtual images in a mirror placed along the imaginary axis. 
For a rotating black hole, this symmetry is broken, and the mirror modes are no longer located at symmetric positions. 
Nevertheless, the term ``mirror modes'' is conventionally retained for the rotating case. 
The mirror modes can sometimes play an important role~\cite{Berti:2005ys,Lim:2019xrb,Dhani:2020nik,Finch:2021iip,Oshita:2024wgt}. 

Throughout this paper, we adopt a convention where positive overtone indices $n = +0, +1, +2, \cdots$ correspond to ordinary modes, and negative indices $n = -0, -1, -2, \cdots$ to mirror modes.
Note that we explicitly include $\pm 0$ to distinguish the original fundamental mode from its mirror counterpart. 

It is also common to classify QNMs into ``prograde'' (or ``corotating'') and ``retrograde'' (or ``counterrotating'') modes, defined by $\mathrm{sign}(\Re(\omega_{lmn}))=\mathrm{sign}(m)$ and $\mathrm{sign}(\Re(\omega_{lmn}))=-\mathrm{sign}(m)$, respectively. 
This classification reflects the behavior of the field phase, which is expressed as $e^{-i(\mathrm{Re}(\omega)t - m\varphi)}$. 
For prograde modes, surfaces of constant phase move in the direction of increasing azimuthal angle, whereas for retrograde modes they move in the opposite direction. This classification does not apply to modes with $m = 0$. In our analysis, we focus on modes with $m > 0$, for which the original/mirror distinction coincides with the prograde/retrograde classification.

There exists a useful symmetry relation between the QNM frequencies and excitation factors of the original and mirror modes.
The homogeneous radial Teukolsky equation is invariant under the combined operation of complex conjugation and the replacements $m \to -m$ and $\omega \to -\omega^*$.
This symmetry implies that the mirror mode frequency is related to the ordinary mode frequency via
\begin{align}
    \omega_{lm-n} = -\omega^*_{l-mn}.
    \label{eq:mirrormode}
\end{align}
Using this relation, we can derive an alternative expression for the contribution of the mirror modes in the ringdown. 
The derivation of this expression, together with a resolution of some confusion in the literature, is summarized in Appendix~\ref{appendix:mirror}.

Since the same symmetry applies to the radial Teukolsky solutions, the ``in'' solution satisfies $R^\text{in*}_{lm}(\omega) = R^\text{in}_{l-m}(-\omega^*)$. 
Consequently, the asymptotic amplitudes also satisfy a similar relation, $B^\text{ref/inc*}_{lm}(\omega) = B^\text{ref/inc}_{l-m}(-\omega^*)$. 
Using this relation with Eqs.~\eqref{eq:excitationfactor} and \eqref{eq:mirrormode}, the excitation factor of the mirror mode is related to that of the ordinary mode via~\footnote{
Equation~\eqref{eq:excitationfactorofmirrormode} exactly matches Eq.~(39) of Ref.~\cite{Lo:2025njp} aside from notation, while it is inconsistent with Eq.~(7) of Ref.~\cite{Oshita:2025ibu}, which can only be obtained by assuming the opposite sign on the right-hand side of Eq.~\eqref{eq:excitationfactorofmirrormode}.}
\begin{align}
    B_{lm-n} = B^*_{l-mn}.
    \label{eq:excitationfactorofmirrormode}
\end{align}

Figure~\ref{fig:qnmfreqandexcitationfactor} shows the QNM frequencies and excitation factors for $(l,m) = (2,2)$ and $(l,m) = (3,1)$ up to the seventh overtone.
Throughout this paper, as a reference for the QNM frequencies and excitation factors, we adopt a publicly available dataset~\cite{Motohashi:2024fwt,motohashi_2024_12696858}, hereafter referred to as the reference dataset (RD).
For $(l,m) = (2,2)$, we highlight the region with the spin parameter $0.85\leq a/M\leq 0.95$ using thick curves where the avoided crossing and resonant excitation occur.
As seen in Fig.~\ref{fig:qnmfreqandexcitationfactor}, the excitation factors of the fifth and sixth overtones exhibit a significant enhancement around $a/M = 0.9$, where the corresponding QNM frequencies approach each other. 
This resonant excitation phenomenon universally occurs in the QNMs in the vicinity of exceptional points~\cite{Motohashi:2024fwt}.
Near this spin value, the excitation factors exhibit a distinctive knot-shaped loop, serving as a unique tool to characterize the underlying spacetime structure.

The $(l,m) = (3,1)$ modes show a complex structure characterized by multiple avoided crossings and resonances~\cite{Motohashi:2024fwt}.
Within this sequence of successive resonances, the resonance between $n=+5$ and $+6$ in the spin range $0.972 \leq a/M \leq 0.99725$ stands out as a sharp resonance.
In Fig.~\ref{fig:qnmfreqandexcitationfactor}, in addition to this sharp resonance, we also highlight the range $0.952 \leq a/M \leq 0.96$ for the mild resonance between $n=+4$ and $+5$ as one of the successive resonances.
We shall discuss the role of the mild resonance in the extraction procedure in Sec.~\ref{sec:results31}.

Finally, we highlight an important property of the excitation factors in the near-extremal regime. 
In the extremal limit $a/M\to 1$, QNMs whose frequencies approach the accumulation point ($2M\omega = m$) have excitation factors that tend to zero~\cite{Ferrari:1984zz,Berti:2006wq}, implying that modes with zero damping rates are not excited.
The RD provides high-precision numerical evidence of this behavior at an unprecedented proximity to extremality, up to $a/M=1-10^{-6}$. 
For instance, the $(l,m,n)=(2,2,+6)$ mode, whose frequency approaches the accumulation point, already shows a nearly vanishing excitation factor at this spin value, whereas the $(2,2,+5)$ mode, which does not approach the accumulation point, retains a finite excitation factor.
This evidence strongly suggests that undamped QNMs are not emitted, offering insights into the open problem of the stability of extremal Kerr black holes~\cite{Dafermos:2025int}.
It also suggests that in the extremal limit, only a restricted subset of modes—such as the $(2,2,+5)$ mode—retain finite excitation, while others are not excited.
This selective excitation raises the possibility that such modes play a special role in the ringdown of extremal Kerr black holes, an intriguing direction for future work on their stability and dynamics.

\subsection{\label{sec:remark}Ambiguities in the definition of excitation factors}

The definition of excitation factors is not unique and involves several conventional choices. 
The RD provides, for $0 \leq a/M \leq 1 - 10^{-6}$, QNM frequencies, angular separation constants, and renormalized angular momenta in the Mano-Suzuki-Takasugi (MST) formalism~\cite{Mano:1996gn,Mano:1996vt,Sasaki:2003xr} to 40-digit precision, as well as excitation factors to 16-digit precision.
All these quantities are available for gravitational, electromagnetic, and scalar perturbations, while in this paper we focus on gravitational waves.
All figures presented in this paper that involve QNM frequencies or excitation factors are based on the RD, either directly or for comparison. 

The excitation factors in the RD are defined for the master variable $\Psi_4$ in the Teukolsky formalism with the normalization $2M=1$ and the conventional tortoise coordinate~\eqref{eq:tortoise}. 
Other conventions found in the literature include excitation factors defined for the master variable $X$ in the Sasaki-Nakamura formalism~\cite{SASAKI198185,Sasaki:1981kj,Sasaki:1981sx}, those adopting $M=1$ normalization, those defined for the gravitational-wave strain $h=h_+ -i h_\times$, and/or those adopting alternative choices for the integration constant of the tortoise coordinate.

In particular, the integration constant of the tortoise coordinate introduces an inherent ambiguity in the excitation factors~\cite{Oshita:2024wgt,Lo:2025njp}. 
This makes the choice a subtle but important one, and it can sometimes be a source of confusion in practice.
If we shift the tortoise coordinate by a constant $C$ as $r_*\to r_* + C$, the asymptotic amplitudes in Eq.~\eqref{eq:Rinasymp} are modified as 
\begin{align}
B^\text{ref}_{lm} \to B^\text{ref}_{lm} e^{-i\omega C}, \quad B^\text{inc}_{lm} \to B^\text{inc}_{lm} e^{i\omega C}.
\end{align}
Consequently, the excitation factors~\eqref{eq:excitationfactor} transform as 
\begin{align}\label{eq:eftrans}
B_{lmn} \to B_{lmn}e^{-2i\omega_{lmn} C}.
\end{align}
Note that, since the QNM frequency $\omega_{lmn}$ has a nonzero imaginary part, the shift $C$ affects the amplitude of the excitation factors in addition to their phase.

One might expect that one could make use of this ambiguity to adjust excitation factor values when comparing them to a ringdown waveform.
This is the case when the integration constant $C$ of the tortoise coordinate for a given ringdown waveform is unknown.
However, an underappreciated fact is that, for a theoretically computed waveform with an explicit choice of $C$, one should simply use the same $C$ for the excitation factors, and there is no degree of freedom to introduce an adjustment only to the excitation factors. 
It is clear that, when one introduces a shift $r_*\to r_* + C$ and change the excitation factors $B_{lmn}$ with a factor $e^{-2i\omega_{lmn} C}$ according to Eq.~\eqref{eq:eftrans}, the same factor also appears from $B^\text{ref}_{lm}/B^\text{inc}_{lm}$ in Eq.~\eqref{eq:Psi4atinfinity}, maintaining the exact agreement between Eqs.~\eqref{eq:Psi4atinfinity} and \eqref{eq:ringdownwaveform}.
In other words, since Eq.~\eqref{eq:ringdownwaveform} is directly derived from Eq.~\eqref{eq:Psi4atinfinity} via the residue theorem, such a shift in the integration constant would consistently appear in both equations.
The adjustment by introducing this factor only to the excitation factors is inconsistent.

Related to these ambiguities, earlier works on computing excitation factors reported different values~\cite{Glampedakis:2003dn,Zhang:2013ksa,Oshita:2021iyn}.
Table~\ref{tab:efs} summarizes this comparison, where we do not include the excitation factors provided in Ref.~\cite{Glampedakis:2003dn} since they differ from the RD in a way that could not be traced back to a specific convention.
As pointed out in Ref.~\cite{Motohashi:2024fwt}, the excitation factors obtained in Ref.~\cite{Zhang:2013ksa} included an erroneous factor $e^{i\omega (1 - \sqrt{1 - a^2/M^2})}$ inherited from the earlier version of Ref.~\cite{Sasaki:2003xr}; correcting this yields consistency with the RD.
On the other hand, the excitation factors in Ref.~\cite{Oshita:2021iyn} differ by a factor of $(1-a^2/M^2)^{i\omega}$.
This likely indicates that the tortoise coordinate employed there involved a corresponding shift of the integration constant, $-\tfrac{1}{2}\ln(1-a^2/M^2)$, relative to the conventional one~\eqref{eq:tortoise}---a shift that diverges in the high-spin limit, which limits the reliability of analyses based on the excitation factors of Ref.~\cite{Oshita:2021iyn}, particularly in the high-spin regime.
That shift coincides with the one later introduced in Ref.~\cite{Oshita:2024wgt} to reconcile Eqs.~\eqref{eq:Psi4atinfinity} and \eqref{eq:ringdownwaveform} for the localized source [or more explicitly, Eqs.~\eqref{eq:deltafunctionsource1} and \eqref{eq:deltafunctionsource2} below], interpreted as a ``start time'' of ringdown (see also Sec.~4.2.3 in Ref.~\cite{Berti:2025hly}).

\begin{table*}
    \centering
    \begin{tabular}{cccccc}
        \hline\hline
        & Dataset & Variable & Normalization & Integration constant shift & Method \\
        \hline
        Zhang, Berti, Cardoso \cite{Zhang:2013ksa} & \cite{BertiEFs,CardosoEFs} & $\Psi_4$, $X$ & $2M=1$ & $-\frac{1}{2}(1 - \sqrt{1 - a^2/M^2})$ & Leaver-MST \\
        Oshita \cite{Oshita:2021iyn} & --- & $h$ & $2M=1$  & $-\frac{1}{2}\ln(1-a^2/M^2)$  & HeunG $\Lambda \to 0$ extrapolation \\
        Motohashi (RD) \cite{Motohashi:2024fwt} & \cite{motohashi_2024_12696858} & $\Psi_4$ & $2M=1$ & 0 & Leaver-Nollert-MST \& Norm squared \\
        Lo, Sabani, Cardoso \cite{Lo:2025njp} & \cite{LoEFs,LoPlots} & $\Psi_4$ & $M=1$ & 0 & Sasaki-Nakamura \\
        This work & --- & $\Psi_4$ & $2M=1$ & 0 & HeunC \\
        \hline\hline
    \end{tabular}
    \caption{\label{tab:efs}
    Comparison of excitation factors for gravitational waves in the literature.
    The ``Integration constant shift'' column indicates the shift of the integration constant in the tortoise coordinate relative to the conventional choice in Eq.~\eqref{eq:tortoise}.
    Each nonzero shift in the first and second rows corresponds to a multiplicative factor $e^{i\omega (1 - \sqrt{1 - a^2/M^2})}$ or $(1-a^2/M^2)^{i\omega}$ relative to the RD.
    }
\end{table*}

In contrast, the RD was recently computed with high precision and established using two independent methods: an improved Leaver-Nollert-MST calculation and a formula based on the norm squared, defined via a regularized biorthogonal product~\cite{Motohashi:2024fwt}.
This marks the first time the excitation factors were rigorously verified through multiple independent  methods.
Further, the RD was later corroborated by a third independent calculation based on the Sasaki-Nakamura formalism~\cite{Lo:2025njp}, which are also publicly available~\cite{LoEFs} along with interactive plots of the QNM frequencies and excitation factors~\cite{LoPlots}.

Here, we provide yet another consistency check of the RD using the Black Hole Perturbation Toolkit~\cite{BHPToolkit}.
Specifically, we compute the excitation factors~\eqref{eq:excitationfactor} by numerically evaluating the asymptotic amplitudes using \texttt{TeukolskyRadial} function in \texttt{Teukolsky} package, employing either \texttt{MST} or \texttt{HeunC} method.
Both methods yield mutually consistent results and show excellent agreement with the RD.
Together with the previous three calculations, this provides four independent and mutually consistent validations, firmly establishing the RD as a robust reference for studies of QNMs.

In summary, the definition of excitation factors involves several conventional choices, including the normalization, the master variable, and in particular the integration constant $C$ of the tortoise coordinate. 
Among these, the tortoise coordinate choice is the most prone to confusion, and using excitation factors and waveforms with different values of $C$ leads to an apparent inconsistency. 
In this paper, we adopt the conventional tortoise coordinate~\eqref{eq:tortoise} employed in the RD.
With this setup, we find that Eqs.~\eqref{eq:Psi4atinfinity} and \eqref{eq:ringdownwaveform} induced by a localized source [see Eqs.~\eqref{eq:deltafunctionsource1} and \eqref{eq:deltafunctionsource2} below] agree without the need for any additional factor. 
We shall confirm this agreement numerically in \S\ref{sec:results3}.

\subsection{\label{sec:waveform}Waveform}

With the general theory of ringdown waveform and excitation factor established, we can now discuss a specific waveform model to study.
Conventionally, QNM fitting has been studied using three main approaches:
(i) fitting synthetic waveforms constructed as a finite sum of damped sinusoids,
(ii) fitting numerically generated gravitational waveforms from perturbation or numerical relativity simulations, and
(iii) fitting observed gravitational wave data.
Ultimately, the goal is to apply approach (iii) to extract QNM frequencies and amplitudes from observational data for black hole spectroscopy.
However, observational noise makes this challenging, so approaches (i) and (ii) have been used complementarily to test and validate fitting algorithms under controlled conditions.

Approach (i) is particularly useful for verifying whether a fitting method can correctly recover the known input QNMs, since the true QNM parameters are set by hand.
Nevertheless, because the synthetic waveform consists of only a finite number of damped sinusoids, it does not capture the full structure of a realistic ringdown, which includes an infinite sum of QNMs plus a power-law tail.
Approach (ii), in contrast, provides waveforms that include all QNMs in principle and are free from observational noise, but they still suffer from numerical errors.
Moreover, the QNM amplitudes in such waveforms are not known a priori, since they also include source-dependent contributions that cannot be separated from the excitation factors.
Although recent studies suggest that QNM amplitudes are largely determined by excitation factors as mentioned above, there is no straightforward way to directly validate the amplitudes obtained through extraction.

We address this gap and benchmark the iterative QNM extraction by considering a Kerr ringdown waveform generated by localized initial data, modeled as a delta-function source term~\cite{Andersson:1996cm,Berti:2006wq}.
The ringdown waveform is obtained by integration in the frequency domain [see Eq.~\eqref{eq:deltafunctionsource1} below], rather than evolving the initial data.
In this special case, the QNM amplitudes are given solely by the excitation factors. 
Combined with the crossvalidated excitation factors from the RD, our analysis enables a direct and reliable comparison of the time-domain ringdown waveform with the frequency-domain resonant excitation.

Hereafter, we focus on a contribution from single $(l,m)$ multipole.
Let us consider the gravitational waveform generated by a localized source.
Specifically, we adopt the following delta-function source term~\cite{Andersson:1996cm,Berti:2006wq}
\begin{align}
    T_{lm}(r') = \frac{r^2 e^{-2i\omega r_*}}{2\pi} \delta(r'-r).
\end{align}
Substituting this source term into Eq.~\eqref{eq:sourcedependnetterm}, we have
\begin{align}
    I_{lm} = -r e^{-i\omega r_*},
\end{align}
where $r$ (or equivalently $r_*$) specifies the fixed location of the source, rather than an argument of $I_{lm}$.
Neglecting the angular dependence, i.e., the factor $S_{lm}e^{im\varphi}/\sqrt{2\pi}$, the $(l,m)$ component of Eq.~\eqref{eq:Psi4atinfinity} reduces to 
\begin{align}
    \Psi_{4,lm}(t) = \frac{1}{2\pi}\int_{-\infty}^{\infty}\mathrm{d}\omega e^{-i\omega t} \frac{1}{2i \omega}\frac{B_{lm}^\text{ref}}{B_{lm}^\text{inc}}.
    \label{eq:deltafunctionsource1}
\end{align}

\begin{figure}
    \includegraphics[width=1.0\linewidth]{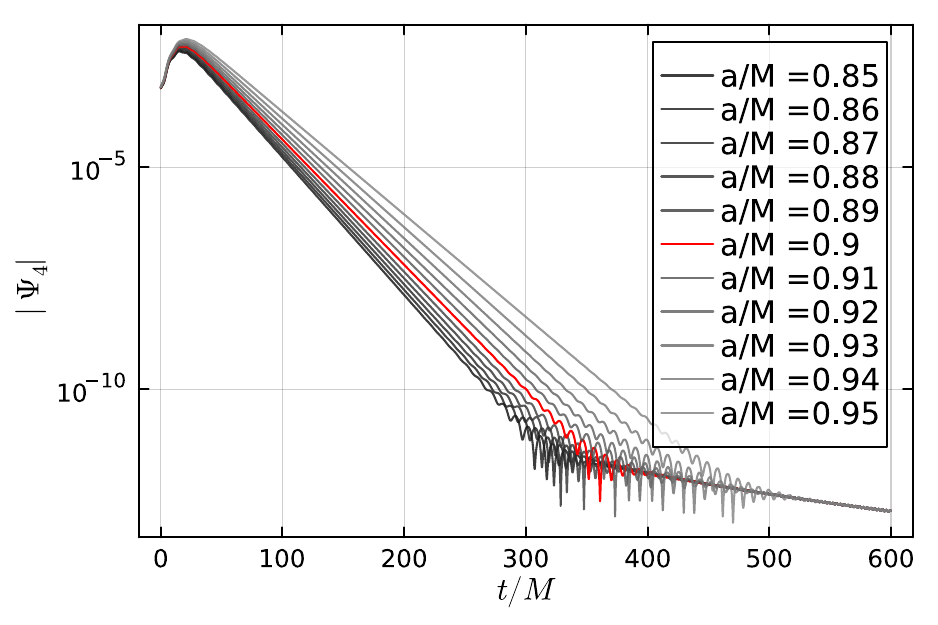}
    \caption{\label{fig:waveform_localized_source} Time-domain ringdown waveforms of $(l,m)=(2,2)$ mode induced by a localized source. The waveform can be found in Ref.~\cite{kubota_2026_18511201}.}
\end{figure}

To numerically evaluate the integral~\eqref{eq:deltafunctionsource1}, we employ the \texttt{Teukolsky} package in the Black Hole Perturbation Toolkit~\cite{BHPToolkit} to compute the asymptotic amplitudes $B^\text{inc}_{lm}$ and $B^\text{ref}_{lm}$. 
These coefficients are also available via the \texttt{GeneralizedSasakiNakamura} package~\cite{Lo:2023fvv}. 
In Fig.~\ref{fig:waveform_localized_source}, we show the ringdown waveforms of $(l,m)=(2,2)$ mode obtained by numerically evaluating Eq.~\eqref{eq:deltafunctionsource1} for the spin parameter $0.85\leq a/M\leq 0.95$.

On the other hand, we can analytically calculate the same integral~\eqref{eq:deltafunctionsource1} using the residue theorem, yielding the following expression:~\footnote{Note that the clockwise integration contour introduces a negative sign in the first term.}
\begin{align}
    \Psi_{4,lm}(t) = - \sum_{n=-\infty}^{\infty} B_{lmn} e^{-i\omega_{lmn} t} + \text{(power-law\ tail)}.
    \label{eq:deltafunctionsource2}
\end{align}
The expression~\eqref{eq:deltafunctionsource2} shows that $\Psi_4$ can be expressed as a superposition of an infinite number of QNMs plus power-law tails.
Importantly, the QNM amplitudes in this expression are precisely equal to the excitation factors $B_{lmn}$.

In summary, we compute the waveform by numerically integrating along the real frequency axis in Eq.~\eqref{eq:deltafunctionsource1}, and then extract the QNM amplitudes from the resulting time-domain waveform using the fitting method described in Sec.~\ref{sec:method}.
With Eq.~\eqref{eq:deltafunctionsource2}, these amplitudes are directly compared with the excitation factors from the RD.
This setup provides a controlled and idealized environment for testing the extraction of QNM amplitudes from a waveform composed of an infinite sum of QNMs and power-law tails.

Moreover, this setup serves not only as a theoretical testbed, but also as a simplified model for realistic waveforms, such as those produced in binary black hole mergers.
As mentioned in Sec.~\ref{sec:intro}, the dependence of QNM amplitudes on the overtone index $n$ in such waveforms is primarily determined by the excitation factors.  
Therefore, although idealized, the waveform employed here captures essential features of realistic gravitational wave signals.  
The insights gained from this analysis can thus be applied directly to the fitting of observational waveforms.

\section{\label{sec:method}Fitting method}
In this section, we describe the fitting procedure to extract QNM amplitudes from the ringdown waveform.  
Our method consists of two main steps: (i) fitting and subtracting the leading power-law tail, and (ii) decomposing the residual signal into a superposition of QNMs (including mirror modes) using an iterative fitting algorithm~\cite{Takahashi:2023tkb}.
We describe the treatment of the tail and the QNM fitting process in detail below.
For notational simplicity, we omit the indices $l,m$ in this section.

\subsection{Power-law tail}
While the ringdown contains a superposition of an infinite number of power-law tails, we focus on the leading contribution.
In principle, we could fit the tail using a fully nonlinear fitting procedure. 
However, in general, nonlinear fitting is often numerically unstable compared to linear fitting. 
To address this, we reduce the problem to considering a single-tail term
\begin{align}
    \Psi_\mathrm{tail}(t) = \frac{A}{(t - t_0)^p},
\end{align}
characterized by complex amplitude $A$, power-law index $p$, and starting time $t_0$.
Taking the logarithm, we have
\begin{align}
    \log \Psi_\mathrm{tail} = \log A + p\log(t-t_0).
\end{align}
This reduces the problem to a linear regression for $\log A$ and $p$ for a given nonlinear parameter $t_0$.

We perform a linear least-squares fit to $\log A$ and $p$ for a range of candidate $t_0$ values, calculating the sum of squared residuals for each trial.
The optimal $t_0$ is chosen as the value that minimizes the sum of squared residuals, and the corresponding $\log A$ and $p$ values are adopted as the best-fit parameters.

\subsection{QNM}
After subtracting the leading tail contribution, we fit the remaining signal with a superposition of QNMs using the iterative fitting method~\cite{Takahashi:2023tkb}. We improve the original study by including mirror modes. While mirror modes are small for spin-aligned, nearly equal-mass binary black hole mergers, which was the primary focus of Ref.~\cite{Takahashi:2023tkb}, they become significant in the asymmetric mass ratio cases~\cite{Cheung:2023vki}, which we investigate here.

\subsubsection{Fitting algorithm at each iteration}
Let us first explain the fitting algorithm at each iteration. 
We employ the fitting function constructed by a superposition of QNMs
\begin{align}
    \psi^\text{fit}(t) = \sum_{n\in \{\mathrm{QNM}\}}C_n \psi_{n}(t),
    \label{eq:fittingfunction}
\end{align}
where \{QNM\} denotes the set of included modes, $\psi_{n}(t)=e^{-i\omega_n t}$ represents the damped sinusoid with $n$-th QNM frequency, and $C_n$ is complex mode amplitude.
The set $\{C_n\}$ serves as fitting parameters.

We introduce the overlap $\rho$ between the waveform $\Psi$ and the fitting function $\psi^\text{fit}$
\begin{align}
    \rho^2 = \frac{|\braket{\psi^\text{fit}}{\Psi}|^2}{\braket{\Psi}{\Psi}\braket{\psi^\text{fit}}{\psi^\text{fit}}},
\end{align}
with
\begin{align}
    \braket{\psi}{\phi}=\int_{t_\text{i}}^{t_\text{e}}\psi^*\phi\mathrm{d}t,
\end{align}
where $t_\text{i}$ and $t_\text{e}$ denote the start and end times of the fitting interval.

We rewrite the overlap using the notation $E_n=\braket{\phi_n}{\Psi}$ and $D_{nm}=\braket{\phi_n}{\phi_m}=(e^{i(\omega_n^*-\omega_m)t_\text{e}}-e^{i(\omega_n^*-\omega_m)t_\text{i}})/(i(\omega_n^*-\omega_m))$ as
\begin{align}
    \rho^2 = \frac{|\sum_n C_n^* E_n|^2}{\braket{\Psi}{\Psi}\sum_{n,m}C^*_n D_{nm} C_m}.
\end{align}
The coefficient $C_n$ is determined by minimizing the overlap as~\cite{Cook:2020otn}
\begin{align}
    C_{n}= (D^{-1})_{nm} E_m.
\end{align}
The maximum overlap $\rho_\text{max}^2$ is then given by
\begin{align}
    \rho_\text{max}^2 = \frac{\sum_{nm}E^*_n(D^{-1})_{nm}E_m}{\braket{\Psi}{\Psi}}
\end{align}
We define the mismatch as
\begin{align}
    \mathcal{M}=1-\rho_\mathrm{max},
\end{align}
which serves as a measure of the goodness of fit.

\subsubsection{Iterative fit}
We iteratively perform the fitting described above, extracting QNM amplitudes, basically in order of their damping times. 
We first determine the amplitude of the longest-lived mode, $C_{+0}$, in the following way. 
We employ the fitting function $\psi^\text{fit}(t)$ with the QNM set shown in the first row of Table \ref{tab:QNMset} and extract the amplitude of the longest-lived mode $C_{+0}$. 
We set the end time $t_\mathrm{e}$ of the fit near the time when the mode we extract falls below the noise level or the next-leading tail component. 

To ensure a stable extraction, we check the variation of $C_n$ with respect to the start time $t_i$, using~\cite{Takahashi:2023tkb}
\begin{align}
    \gamma_n = \left|\frac{1}{C_n}\frac{\mathrm{d}C_{n}}{\mathrm{d} t_i}\right|,
\end{align}
which measures the flatness of the plateau of the mode amplitude $C_n$.
We select the start time $t_i$ of the fitting interval at which $\gamma_{+0}$ reaches its minimum value.
Using this $t_i$, we extract the mode amplitude $C_{+0}$.

Next, we determine the amplitude of the next-longest-lived mode, $C_{-0}$. 
We subtract the previously extracted contribution, $\Psi(t)-C_{+0}e^{-i\omega_{+0}(t)}$, and fit the residual waveform using the QNM set in the second row of Table~\ref{tab:QNMset}.
We then extract $C_{-0}$, which is now the effective longest-lived mode in the residual, by the same flatness criterion, adopting the start time $t_\text{i}$ that minimizes $\gamma_{-0}$.
This process is then iteratively repeated for subsequent modes, basically following the order of their lifetimes. 

However, we find that, when mode amplitudes are small, it is more efficient to adjust the extraction order based on the amplitude hierarchy.
In this study, since $|C_{-n}|$ are small for $n\geq 1$, we skip them and instead proceed to extract the prograde higher overtones.
The QNM sets employed at each iteration are summarized in Table~\ref{tab:QNMset}, where the ``Extract'' column indicates the mode determined at each step.

Note that each step carries numerical uncertainties, which propagate to subsequent steps.
These numerical uncertainties contaminate the residual waveform and may eventually affect the extracted amplitudes, in particular those of higher overtones.
Such contamination should be taken into account in addition to the stability of the extraction procedure, which we also address explicitly below.

\begin{table}[]
    \begin{tabular}{ccc}
    \hline\hline
        Step & Extract & \{QNM\} \\ \hline
        $1$ & $+0$ & $\{+0, +1, +2, +3, -0, -1, -2, -3\}$ \\
        $2$ & $-0$ & $\{+1, +2, +3, -0, -1, -2, -3\}$\\
        $3$ & $+1$ & $\{+1, +2, +3, +4, -1, -2, -3\}$\\
        $4$ & $+2$ & $\{+2, +3, +4, +5, -1, -2, -3\}$\\
        $5$ & $+3$ & $\{+3, +4, +5, +6, -1, -2, -3\}$\\
        $6$ & $+4$ & $\{+4, +5, +6, +7, -1, -2, -3\}$\\
        $7$ & $-1$ & $\{+5, +6, +7, -1, -2, -3\}$\\
        $8$ & $+5$ & $\{+5, +6, +7, -2, -3\}$\\
        $9$ & $+6$ & $\{+6, +7, -2, -3\}$\\
        $10$ & $+7$ & $\{+7, -2, -3\}$\\
    \hline\hline
    \end{tabular}
    \caption{\label{tab:QNMset} Progression of extracted $(l,m)=(2,2)$ overtones and the corresponding QNM sets at each step. The ``Extract'' column indicates the overtone index of the QNM extracted in each step, while the ``\{QNM\}'' column lists the set of QNMs included in the fitting function at each step.}
\end{table}

\section{\label{sec:results}Results}
In this section, we present the results of the iterative QNM fitting procedure with mirror modes described in \S\ref{sec:method}.  
We analyze waveforms generated by localized initial data given in Eq.~\eqref{eq:deltafunctionsource1} and assess the stability and accuracy of the extracted QNM amplitudes by comparing them with the excitation factors from the RD.
Below, we focus on two representative cases of resonant excitation in Kerr QNMs~\cite{Motohashi:2024fwt}: 
a mild resonance in $(l,m)=(2,2)$ at $a/M\simeq 0.9$, and a sharp resonance in $(l,m)=(3,1)$ at $a/M\simeq 0.9722$ occurring as part of a sequence of successive resonances.
In addition to the fitting analysis, we also explore iterative subtraction of exact QNMs, which serves as a benchmark mimicking an ideal iterative fitting.
Our analysis reveals characteristic contributions from resonance modes and elucidates their distinctive features.

\subsection{Stability of mode extraction}
\begin{figure}
    \includegraphics[width=\columnwidth, page=51]{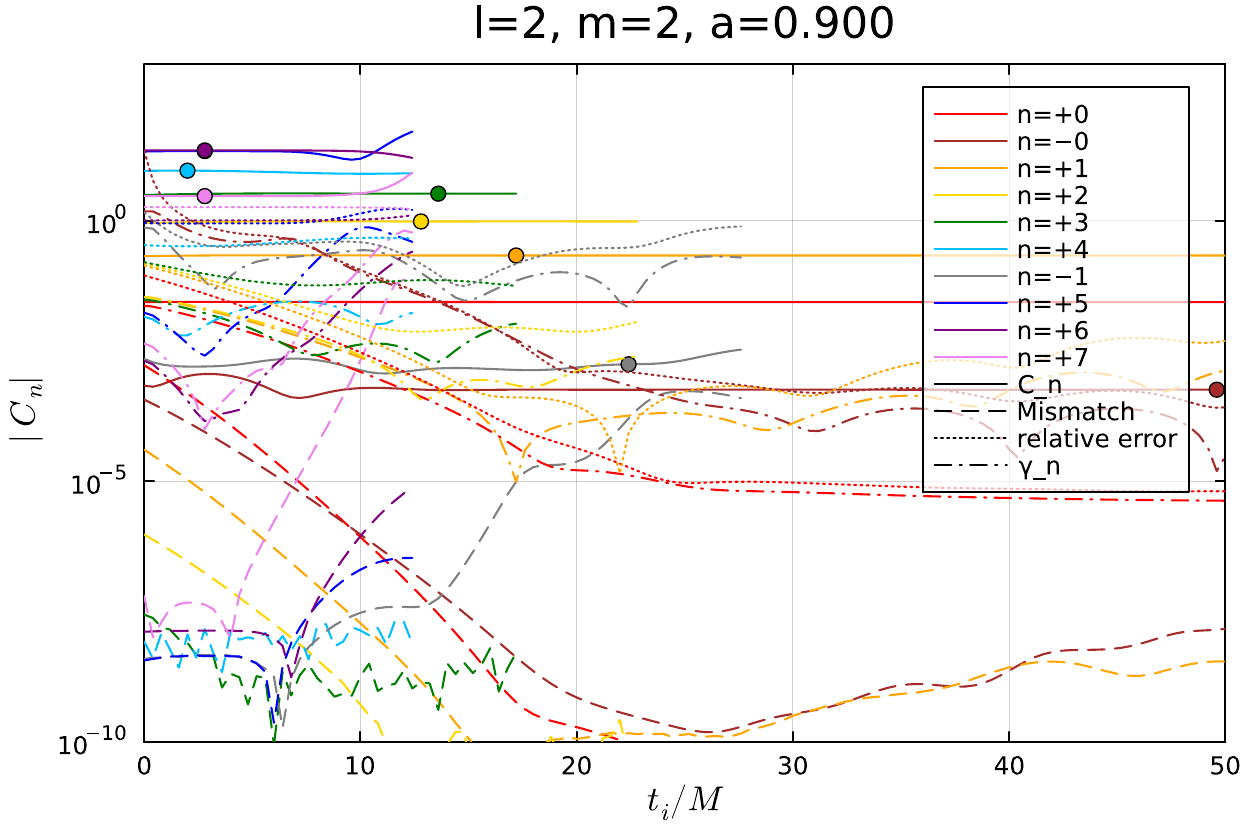}
    \includegraphics[width=\columnwidth, page=51]{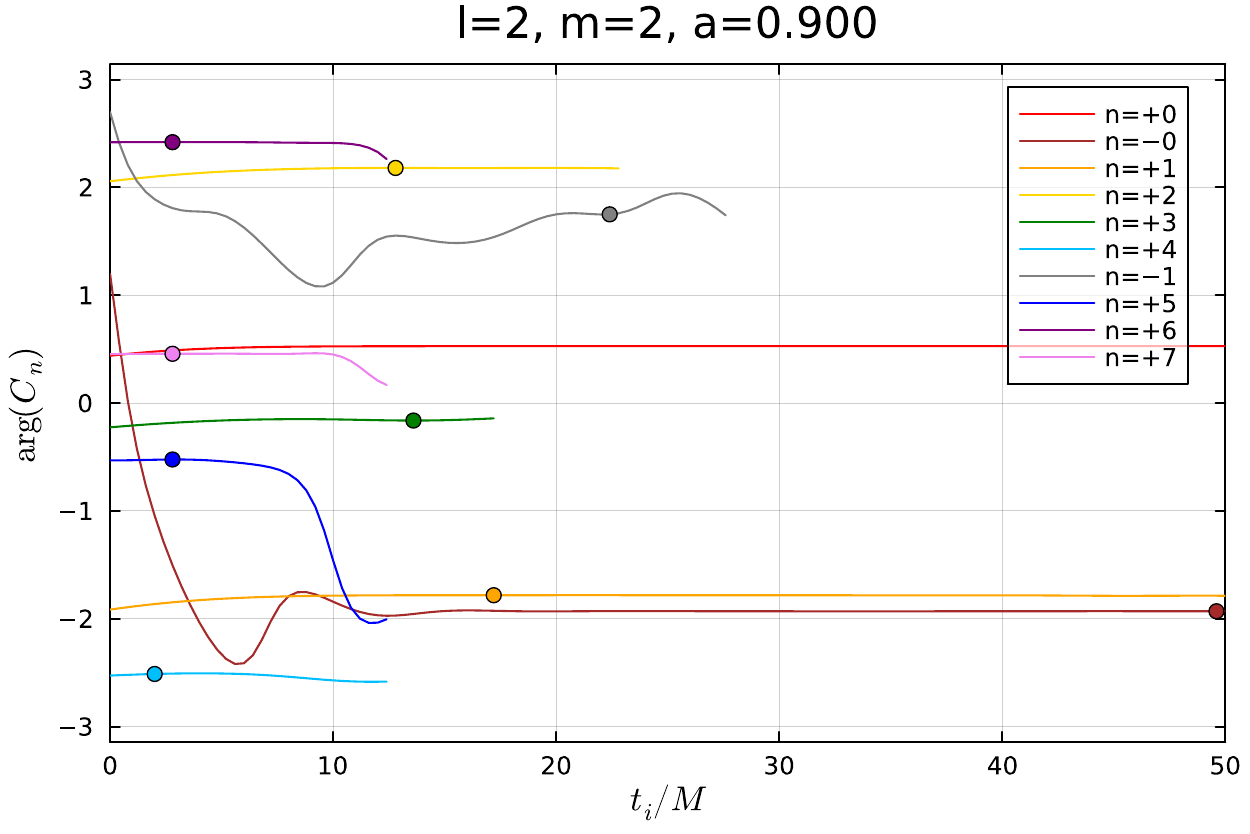}
    \caption{\label{fig:Stability}Stability of QNM amplitude extraction for $(l,m)=(2,2)$ at $a/M=0.9$ with respect to variations of the fitting start time $t_i$. The top panel shows the extracted amplitude (solid line), the mismatch (dashed line), and the relative error between the extracted amplitude and excitation factor (dotted line). Circular markers indicate the extracted amplitudes and their corresponding extraction times, where $\gamma_n$ is minimized. The bottom panel shows the phase of the extracted amplitude in the complex plane.}
\end{figure}

Figure~\ref{fig:Stability} illustrates the stability of the mode extraction for $(l,m)=(2,2)$ at $a/M=0.9$. 
In the top panel, the solid line shows the extracted amplitude as a function of the starting time for the fitting interval, while the dashed line indicates the mismatch. 
The dotted line represents the relative error between the extracted amplitude and the excitation factor.
Circular markers represent the extracted amplitude at the starting time where $\gamma_n$ reaches its minimum.

This analysis demonstrates that we can extract amplitudes within a plateau region, indicating that the extracted amplitude is stable with respect to variations in the starting time. 
For modes up to the third overtone, the relative error remains below $10$\%, confirming that the amplitudes are extracted correctly.
In other words, the excitation factors from the RD are consistent with the amplitudes extracted from the waveforms induced by the localized source.
However, for the fifth and sixth overtones, while the amplitudes are extracted in a stable (plateau) region, they are accurate only to within $\mathcal{O}(1)$ errors.

\subsection{Residual waveform and contamination}
\begin{figure*}
    \begin{minipage}[]{0.49\textwidth}
        \includegraphics[width=\columnwidth, page=1]{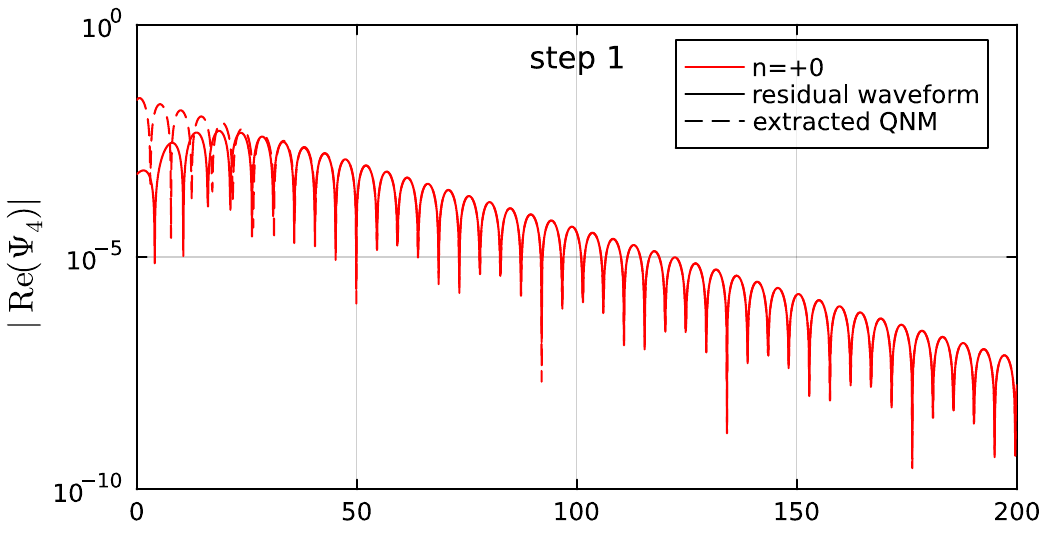}
        \vfill
        \includegraphics[width=\columnwidth, page=2]{figure/real_waveform_a090_fit.pdf}
        \vfill
        \includegraphics[width=\columnwidth, page=3]{figure/real_waveform_a090_fit.pdf}
        \vfill
        \includegraphics[width=\columnwidth, page=4]{figure/real_waveform_a090_fit.pdf}
        \vfill
        \includegraphics[width=\columnwidth, page=5]{figure/real_waveform_a090_fit.pdf}
    \end{minipage}
    \begin{minipage}[]{0.49\textwidth}
        \includegraphics[width=\columnwidth, page=6]{figure/real_waveform_a090_fit.pdf}
        \vfill
        \includegraphics[width=\columnwidth, page=7]{figure/real_waveform_a090_fit.pdf}
        \vfill
        \includegraphics[width=\columnwidth, page=8]{figure/real_waveform_a090_fit.pdf}
        \vfill
        \includegraphics[width=\columnwidth, page=9]{figure/real_waveform_a090_fit.pdf}
        \vfill
        \includegraphics[width=\columnwidth, page=10]{figure/real_waveform_a090_fit.pdf}
    \end{minipage}
    \caption{\label{fig:snapshowofextraction}Snapshots of iterative extraction from the waveform for $(l,m)=(2,2)$ at $a/M=0.9$. 
    Shown are the residual waveform (solid line) after subtracting the fitted single tail and fitted QNM(s) from the original waveform, and the extracted QNM (dashed line) at each step.
    }
\end{figure*}

Figure~\ref{fig:snapshowofextraction} illustrates the iterative extraction process for $(l,m)=(2,2)$ at $a/M=0.9$. 
Snapshots for steps 1 through 5 are displayed in the left column (top to bottom), while steps 6 through 10 are shown in the right column (top to bottom). 
In each panel, the solid line represents the (residual) waveform to be fitted, and the dashed line denotes the extracted QNM at that step. 
We observe that the damped sinusoid matches the extracted QNM. 
However, after a certain number of steps, the waveform contamination appears.

To diagnose the source of this contamination, we perform an iterative subtraction of exact QNMs from the waveform.
In contrast to the iterative extraction, where the QNM amplitudes are determined by fitting at each step, the iterative subtraction simply removes exact QNM damped sinusoids with amplitudes fixed to the excitation factors from the RD.
This approach enables a case study of how an ideal extraction would behave and provides further diagnostics of the fitting procedure.

\begin{figure*}
    \begin{minipage}[]{0.49\textwidth}
        \includegraphics[width=\columnwidth, page=1]{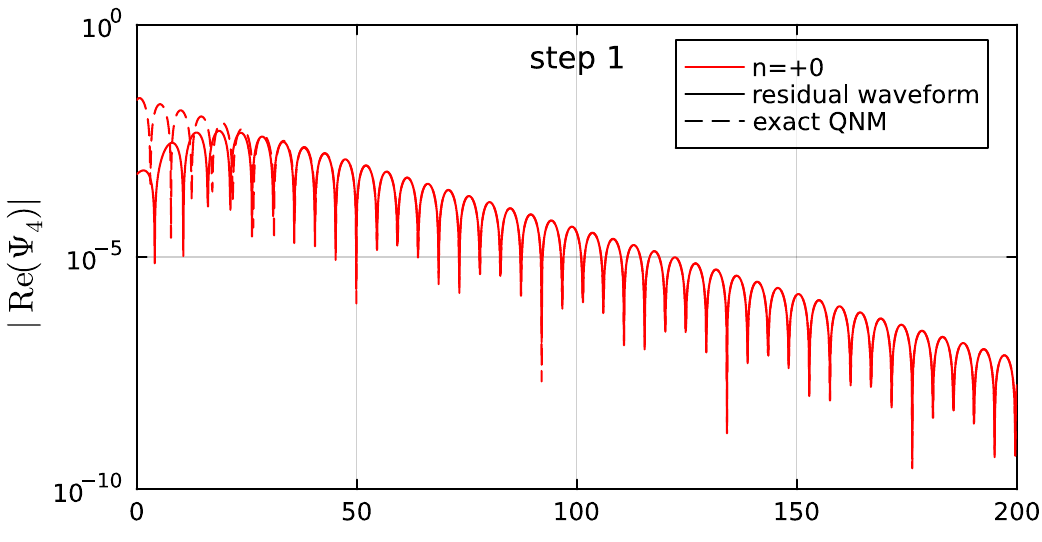}
        \vfill
        \includegraphics[width=\columnwidth, page=2]{figure/real_waveform_a090_exact_qnm.pdf}
        \vfill
        \includegraphics[width=\columnwidth, page=3]{figure/real_waveform_a090_exact_qnm.pdf}
        \vfill
        \includegraphics[width=\columnwidth, page=4]{figure/real_waveform_a090_exact_qnm.pdf}
        \vfill
        \includegraphics[width=\columnwidth, page=5]{figure/real_waveform_a090_exact_qnm.pdf}
    \end{minipage}
    \begin{minipage}[]{0.49\textwidth}
        \includegraphics[width=\columnwidth, page=6]{figure/real_waveform_a090_exact_qnm.pdf}
        \vfill
        \includegraphics[width=\columnwidth, page=7]{figure/real_waveform_a090_exact_qnm.pdf}
        \vfill
        \includegraphics[width=\columnwidth, page=8]{figure/real_waveform_a090_exact_qnm.pdf}
        \vfill
        \includegraphics[width=\columnwidth, page=9]{figure/real_waveform_a090_exact_qnm.pdf}
        \vfill
        \includegraphics[width=\columnwidth, page=10]{figure/real_waveform_a090_exact_qnm.pdf}
    \end{minipage}
    \caption{\label{fig:substractionexactqnm}Snapshots of iterative subtraction of exact QNMs from the waveform for $(l,m)=(2,2)$ at $a/M=0.9$. 
    Line styles are the same as in Fig.~\ref{fig:snapshowofextraction}.
    }
\end{figure*}

In Fig.~\ref{fig:substractionexactqnm}, we show snapshots of iterative subtraction of exact QNMs from the $(l,m)=(2,2)$ waveform.
In this case, no such contamination is observed, and we clearly see the next-leading tail contribution. 
Therefore, we conclude that the contamination observed in Fig.~\ref{fig:snapshowofextraction} originates from errors in the extracted amplitudes during the iterative fitting. 
Such contamination propagates through subsequent steps, limiting the extraction accuracy to approximately the third overtone in our setup, even though higher modes are in principle present.

\begin{figure}
     \begin{tabular}{c}
        \begin{minipage}[]{\linewidth}
            \centering
            \includegraphics[width=1.0\linewidth]{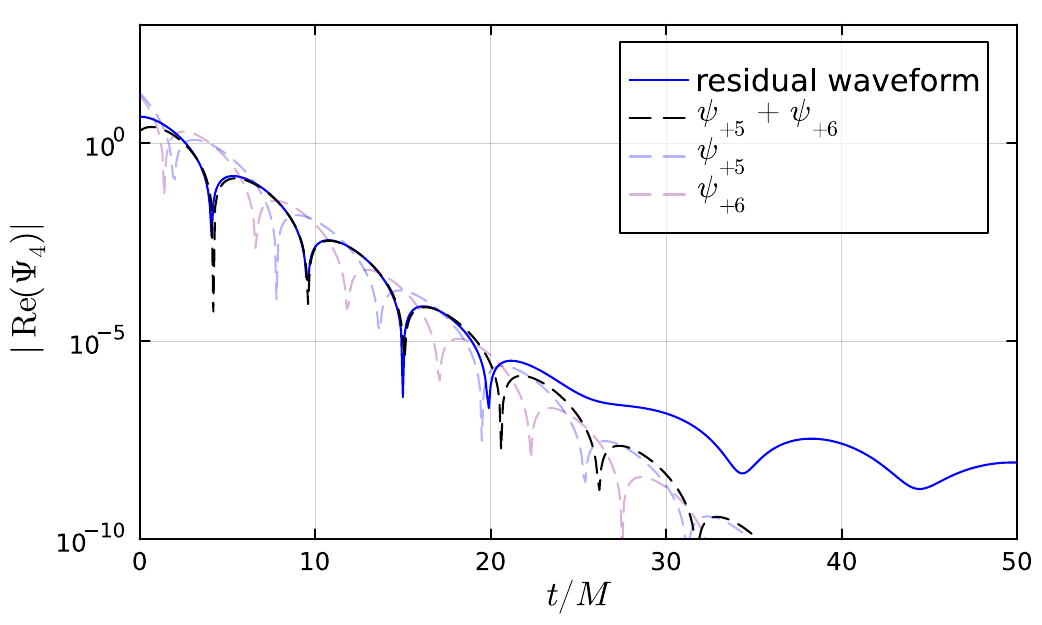}
            \subcaption{Iterative extraction}
        \end{minipage} \\ 
        \begin{minipage}[]{\linewidth}
            \centering
            \includegraphics[width=1.0\linewidth]{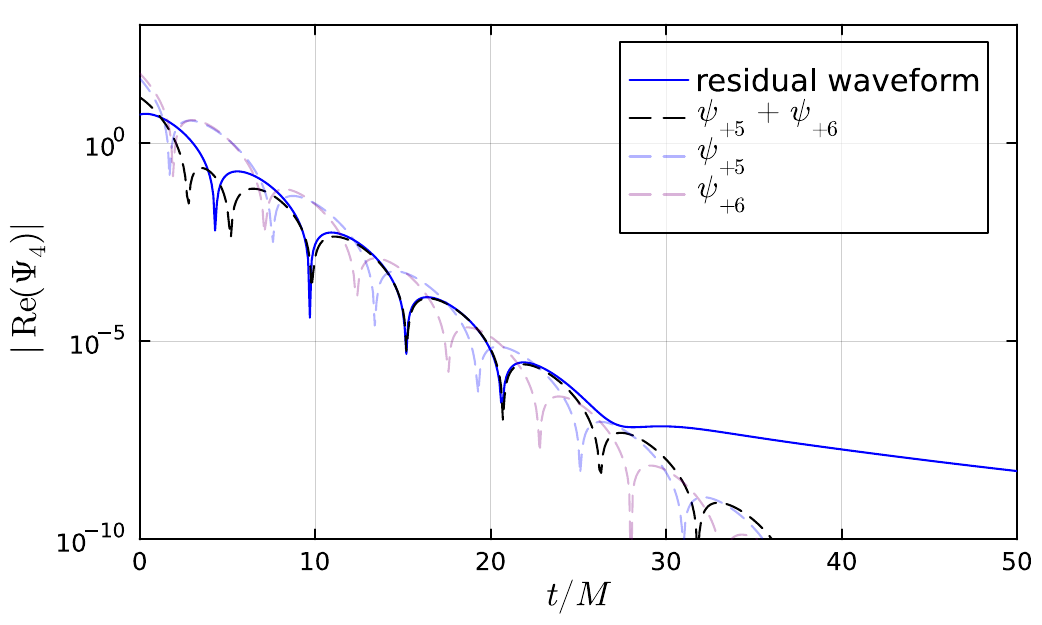}
            \subcaption{Subtraction of exact QNMs}
        \end{minipage} 
    \end{tabular}
    \caption{\label{fig:real_waveform_a0.9-5-6_exactQNM}Residual waveforms at step 8 in Figs.~\ref{fig:snapshowofextraction} and \ref{fig:substractionexactqnm}, compared with the fifth and sixth QNMs and their sum for $(l,m)=(2,2)$ at $a/M=0.9$.}
\end{figure}

It is interesting to see that, in step 8, the extracted QNM (dashed line) does not coincide with the damped sinusoid in the residual waveform (solid line). 
This discrepancy is attributed to the fact that the residual waveform at step 8 is dominated by fifth and sixth QNMs, which have similar damping rates.
While the fifth and sixth QNMs do not individually match the residual damped sinusoid, their combined contribution reproduces the observed oscillation, as shown in Fig.~\ref{fig:real_waveform_a0.9-5-6_exactQNM}.

Nevertheless, we should bear in mind that the extracted amplitudes of the fifth and sixth overtones have $\mathcal{O}(1)$ errors as mentioned above.
We see that the extracted fifth and sixth QNMs appear to be more consistent with the residual waveform at the early time, while the exact fifth and sixth QNMs match the residual waveform at later time.
The early time waveform is actually dominated by higher overtones, so the matching at early regime may signal overfitting.
To extract the fifth and sixth QNMs more accurately, we need longer waveforms with smaller tail contributions and smaller contamination from the lower overtone fitting.

\begin{figure*}
    \includegraphics[width=1.0\textwidth]{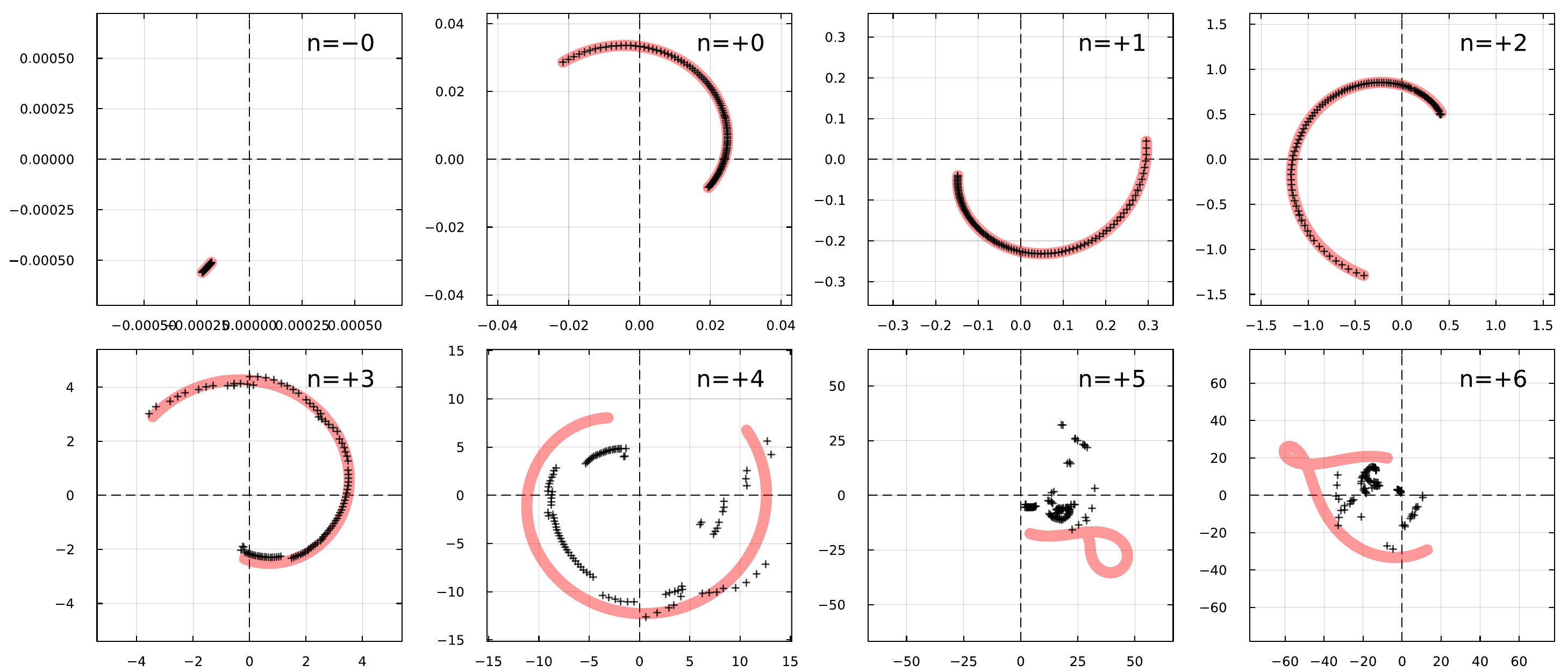}
    \caption{\label{fig:extractedamplitude_deltasource}Extracted amplitudes (black plus markers) from the waveform, compared with excitation factors (red solid lines) for $(l,m)=(2,2)$ at $0.85\leq a/M\leq 0.95$.}
\end{figure*}
\begin{figure*}
    \includegraphics[width=1.0\textwidth]{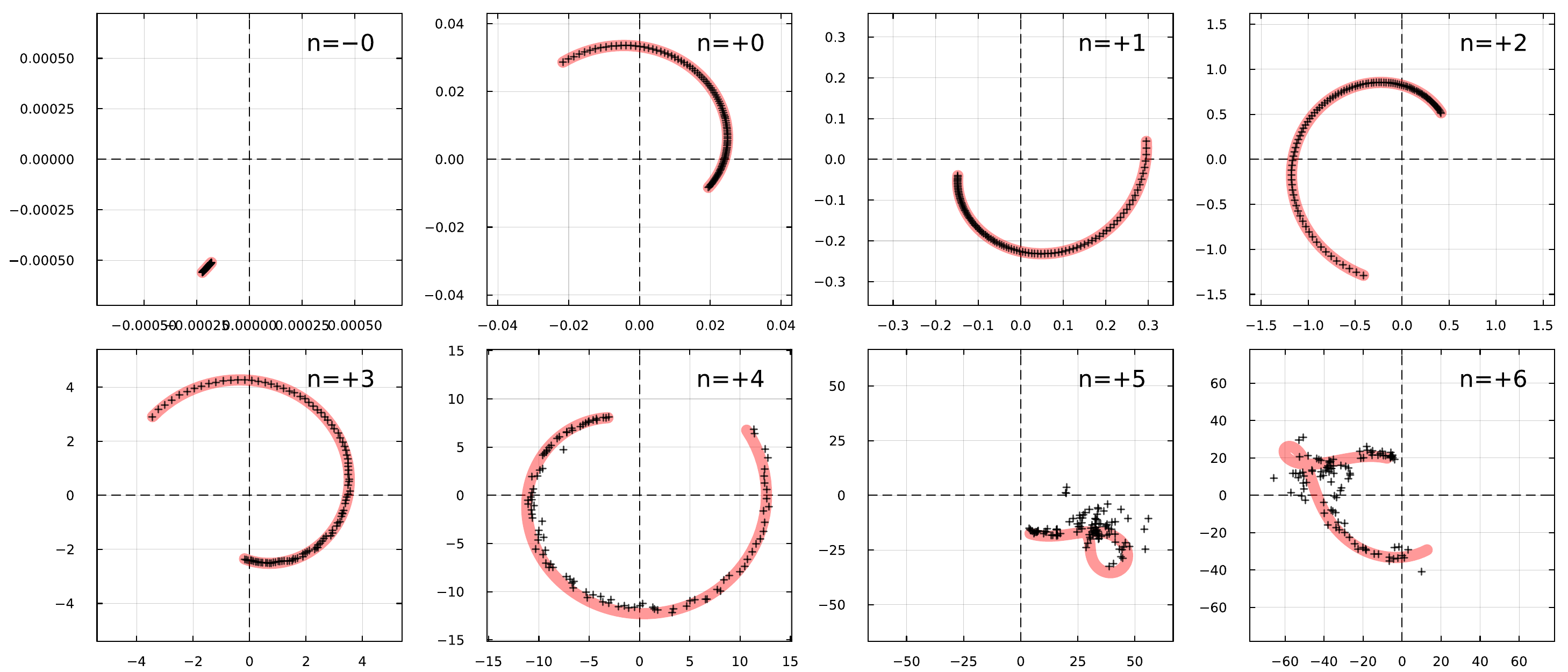}
    \caption{\label{fig:extractedamplitude_deltasource_minimize_relative_error}Extracted amplitudes (black plus markers) at the optimal extraction time determined by minimizing the relative error, compared with excitation factors (red solid lines) for $(l,m)=(2,2)$ at $0.85\leq a/M\leq 0.95$.
    }
\end{figure*}

\subsection{\label{sec:results3}Extracted amplitudes and excitation factors}
Figure~\ref{fig:extractedamplitude_deltasource} shows the extracted amplitudes in the complex plane for $(l,m)=(2,2)$ at $0.85\leq a/M \leq 0.95$. 
Black plus markers represent the extracted amplitudes, while the red solid lines indicate the excitation factors from the RD. 
The black plus markers lie on the red line up to the third overtone, indicating the consistent extraction.
It also provides a numerical validation of the agreement between Eqs.~\eqref{eq:deltafunctionsource1} and \eqref{eq:deltafunctionsource2} without the need for any additional factor.
Deviations begin to appear from the fourth overtone.
As mentioned above, the extracted amplitudes of the fifth and sixth overtones deviate by $\mathcal{O}(1)$ from the excitation factors, and hence the expected resonance structure is not observed.

Figure~\ref{fig:extractedamplitude_deltasource_minimize_relative_error} shows the extracted amplitudes at the fitting start times at which the relative error is minimized. 
The relative error is $\mathcal{O}(10^{-5})$ for $n=+0$, but increases to $\mathcal{O}(10^{-1})$ for $n=+5$ and $n=+6$.

Although we do not yet know how to determine such optimal start times in practice, it is useful to regard them as a benchmark of the ``best fitting.''
In this case, the extracted amplitudes almost lie on the red line. 
The anticipated knot-shaped structure for the fifth and sixth overtones is still absent, but their extracted amplitudes nevertheless display a rough pattern of resonant amplification.
A clearer extraction of resonance structure may require improvements to the fitting algorithm and/or the use of waveforms with smaller tail contributions.

\subsection{\label{sec:results31}Sharp resonances}
Compared to mild resonances, for which the Kerr $(l,m)=(2,2)$ mode is a representative example, sharp resonances occur when two QNM frequencies approach exceptional points more closely. 
In this case, the corresponding excitation factors are strongly amplified, tracing a lemniscate pattern in the complex plane. 
As a representative case of sharp resonances, we investigate the Kerr $(l,m)=(3,1)$ mode. 
Here, avoided crossings and resonances appear successively in multiple overtone pairs, and in particular the pair $n=+5$ and $+6$ exhibits a sharp resonant excitation around $a/M = 0.9722$~\cite{Motohashi:2024fwt}. 
We perform the iterative fitting and find that it leads to broadly similar conclusions as in the mild resonance case, except for the change of overtone hierarchy. 
Below, we focus on highlighting this characteristic feature originating from an intriguing resonance structure.

\begin{figure}
     \begin{tabular}{c}
        \begin{minipage}[]{\linewidth}
            \centering
            \includegraphics[width=1.0\linewidth]{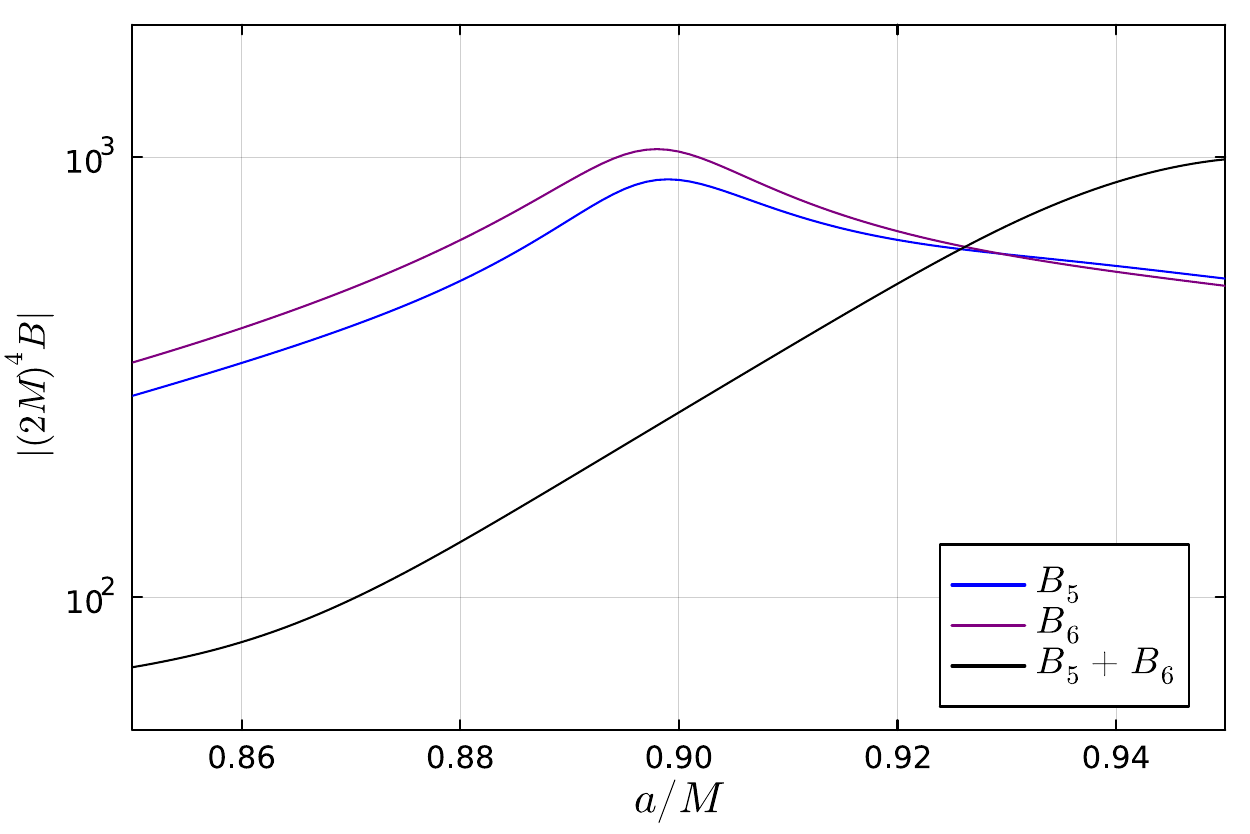}
            \subcaption{$(l,m)=(2,2)$ around $a/M=0.9$}
        \end{minipage} \\ 
        \begin{minipage}[]{\linewidth}
            \centering
            \includegraphics[width=1.0\linewidth]{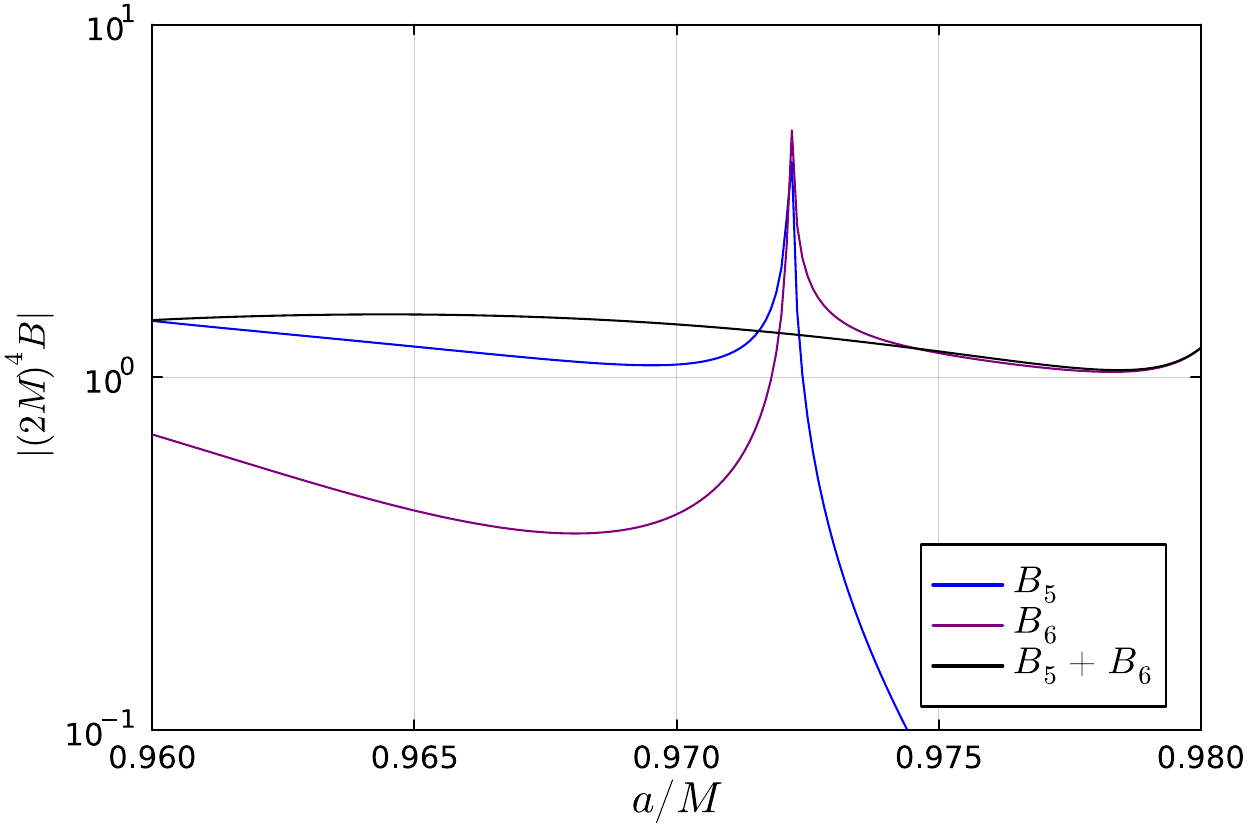}
            \subcaption{$(l,m)=(3,1)$ around $a/M=0.9722$}
        \end{minipage} 
    \end{tabular}
    \caption{\label{fig:abs_excitation} Absolute values of excitation factors for the resonance modes $n=+5$ and $+6$, together with the absolute value of their sum $|B_5+B_6|$.}
\end{figure}

As a simple preliminary check, let us look at the sum of the excitation factors.
In Fig.~\ref{fig:abs_excitation}, we show the resonant pair overtones $n=+5$ and $+6$ in $(l,m)=(2,2)$ and $(3,1)$ modes.
Around the resonant peak of each excitation factor, their sum does not show any dramatic variation.
The sum does not vanish and remains comparable in magnitude, although—as we shall discuss below—this observation alone is insufficient to capture the full impact of resonance on the ringdown waveform.

The stable behavior of the sum is consistent with the fact that the excitation factors follow the lemniscate trajectory in the complex plane~\cite{Motohashi:2024fwt}.
By virtue of the symmetric lemniscate pattern, the sum of the excitation factors during resonance corresponds to the center of the figure-eight shape, and hence does not change significantly.
While the contribution to gravitational-wave strain scales as $\sim \omega^{-2}B$, since $\omega$ does not vary appreciably during the avoided crossing, the above argument applies, especially for sharp avoided crossings and resonances.
The same conclusion holds when introducing an additional parameter controlling the sharpness of the avoided crossing and resonance, as demonstrated in Fig.~S1 of Ref.~\cite{Motohashi:2024fwt}.
This behavior is partially consistent with Ref.~\cite{Oshita:2025ibu}.

However, these qualitative features differ from those shown in Fig.~9 in Ref.~\cite{Oshita:2025ibu}, where the sum of the resonant excitation factors appears to vary significantly, reaching a minimum, and also seems sensitive to the sharpness of the avoided crossing and resonance.
This discrepancy can be traced to a multiplicative factor $(1-a^2/M^2)^{i\omega}$ in the excitation factors (see Table~\ref{tab:efs}).
Notably, the location of the minimum of the summed excitation factors does not coincide with the resonance peak, suggesting irrelevance to any potential cancellation originating from resonance.
We confirm that multiplying this factor, together with $\omega^{-2}$ to match the convention, with the RD excitation factors reproduces both the variation and the apparent cancellation.

If a different integration constant is chosen for the tortoise coordinate, 
effectively corresponding to choosing a different reference time in the ringdown at which the excitation factors are evaluated,
the excitation factors transform as in Eq.~\eqref{eq:eftrans}, and their sum would accordingly exhibit different behavior. 
Therefore, interpreting either the excitation factors themselves or their sum requires special care.
Furthermore, the sum of the resonant excitation factors captures only part of the contributions near exceptional points~\cite{PanossoMacedo:2025xnf} and may even be misleading if considered in isolation. 
For these reasons, the contribution of resonance to the time-domain signal cannot be fully understood by examining only the excitation factors and their sum. 
It is essential to analyze the time-domain waveform directly.

To this end, we apply both iterative extraction of fitted QNMs and iterative subtraction of exact QNMs to the time-domain signal, thus revealing characteristic contributions of sharp resonances in Kerr ringdown. 
As explained above in Figs.~\ref{fig:snapshowofextraction}--\ref{fig:real_waveform_a0.9-5-6_exactQNM}, the contribution from the mild resonance in $(l,m)=(2,2)$ can be seen in the ringdown signal. 
As we shall see below, the sharp resonance together with successive mild resonances in $(l,m)=(3,1)$ exhibit a distinct excitation structure and a nontrivial overtone hierarchy.

\begin{figure*}
    \begin{minipage}[]{0.49\textwidth}
        \includegraphics[width=\columnwidth, page=1]{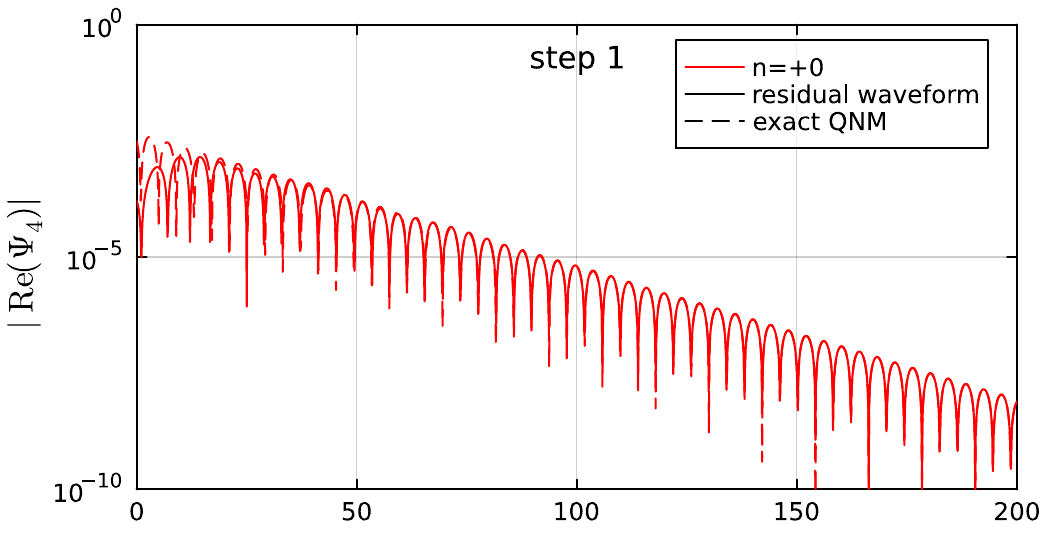}
        \vfill
        \includegraphics[width=\columnwidth, page=2]{figure/real_waveform_31_a09722_exact_qnm.pdf}
        \vfill
        \includegraphics[width=\columnwidth, page=3]{figure/real_waveform_31_a09722_exact_qnm.pdf}
        \vfill
        \includegraphics[width=\columnwidth, page=4]{figure/real_waveform_31_a09722_exact_qnm.pdf}
        \vfill
        \includegraphics[width=\columnwidth, page=5]{figure/real_waveform_31_a09722_exact_qnm.pdf}
    \end{minipage}
    \begin{minipage}[]{0.49\textwidth}
        \includegraphics[width=\columnwidth, page=6]{figure/real_waveform_31_a09722_exact_qnm.pdf}
        \vfill
        \includegraphics[width=\columnwidth, page=7]{figure/real_waveform_31_a09722_exact_qnm.pdf}
        \vfill
        \includegraphics[width=\columnwidth, page=8]{figure/real_waveform_31_a09722_exact_qnm.pdf}
        \vfill
        \includegraphics[width=\columnwidth, page=9]{figure/real_waveform_31_a09722_exact_qnm.pdf}
        \vfill
        \includegraphics[width=\columnwidth, page=10]{figure/real_waveform_31_a09722_exact_qnm.pdf}
    \end{minipage}
    \caption{\label{fig:substractionexactqnm31}Snapshots of iterative subtraction of exact QNMs from the waveform for $(l,m)=(3,1)$ at $a/M=0.9722$. 
    Line styles are the same as in Fig.~\ref{fig:snapshowofextraction}.
    }
\end{figure*}

In Fig.~\ref{fig:substractionexactqnm31}, we show snapshots of the iterative subtraction of exact QNMs from the $(l,m)=(3,1)$ waveform with $a/M=0.9722$, mimicking an ideal iterative fitting procedure. 
As explained above, the sum of the $n=+5$ and $+6$ excitation factors decreases compared to the individual values.
Nevertheless, this reduction is not as severe as to make the combined amplitude extremely small.
In particular, we observe in step 8 in Fig.~\ref{fig:substractionexactqnm31} that the contributions from the $n=+5$ and $+6$ modes exceed that of the $n=+4$ mode by roughly $\mathcal{O}(10^{2-3})$.

In general, aside from fitting error contamination and tail contribution, it has been broadly accepted in the community that, for prograde modes, the lowest (longest-lived) overtone dominates the late-time signal at each step of the iterative procedure, while higher overtones contribute primarily at earlier times.
However, in the present $(l,m)=(3,1)$ case, the $n=+4$ mode is completely buried beneath higher overtones,  whereas the $n=+5$ and $+6$ modes remain clearly visible above the tail. 

The origin of this anomalous behavior lies in the exceptionally small excitation factor of the $n=+4$ mode. 
In the high-spin limit, excitation factors generally tend to approach zero. 
Typically this decrease is nearly uniform, so no extreme hierarchy emerges among the modes. 
Resonances, however, represent precisely the kind of irregular behavior that breaks this trend, producing not only strong \textit{amplification} but also, as in the present case, a pronounced \textit{suppression}.

As shown in Fig.~2 of Ref.~\cite{Motohashi:2024fwt} and confirmed directly from the RD, the $n=+4$ excitation factor approaches zero at a lower spin than the other overtones. 
This behavior results from the mild resonance between $n=+4$ and $n=+5$ at $a/M\simeq 0.956$, which occurs before the sharp resonance between $n=+5$ and $+6$ at $a/M\simeq 0.9722$. 
In general, responses to parameter variations become particularly sensitive in the vicinity of exceptional points.
For $n=+4$, the excitation factor is accelerated by its mild resonance with $n=+5$, leading to a rapid approach toward zero. 

\begin{figure}
    \includegraphics[width=1.0\linewidth]{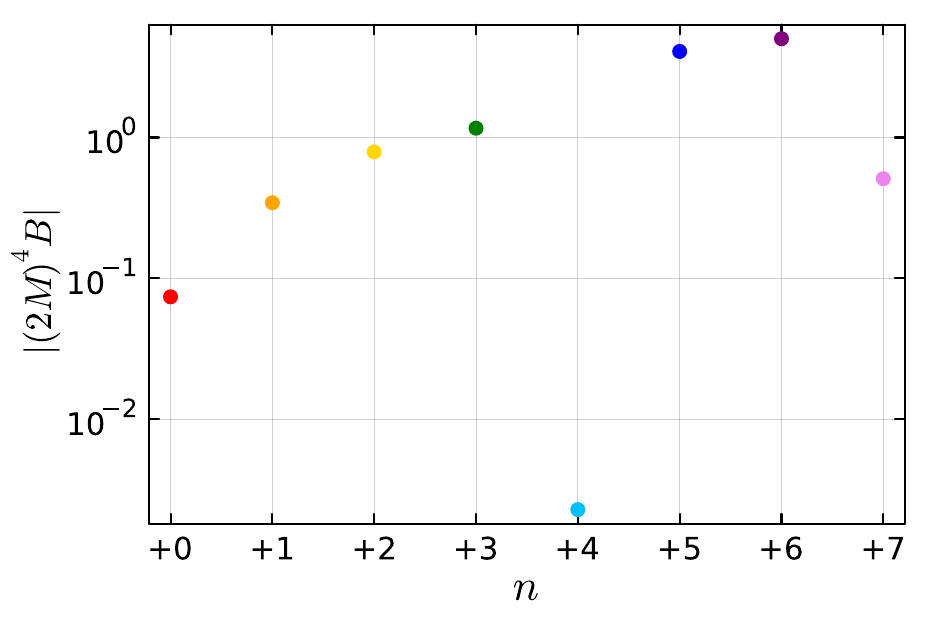}
    \caption{\label{fig:excitation_s-2l3m1a09722}Absolute values of the excitation factors for $(l,m)=(3,1)$ at $a/M=0.9722$.}
\end{figure}

Consequently, $n=+4$ attains an anomalously small value, and within the iterative procedure the contributions from $n=+5$ and $+6$ dominate, overshadowing that of $n=+4$. 
In Fig.~\ref{fig:excitation_s-2l3m1a09722}, we present the absolute values of the excitation factors at $a/M=0.9722$. 
We find $|B_4|\sim 0.002$ and $|B_5|\sim 4$, revealing a separation of three orders of magnitude. 
This hierarchy is directly reflected in their relative contributions to the ringdown signal, as seen in Fig.~\ref{fig:substractionexactqnm31}.

As we discussed in Sec.~\ref{sec:method}, when extracting QNMs, one should prioritize the extraction of the dominant modes first, as this will lead to a more accurate and efficient fitting process. 
Our benchmark analysis shows that, for a waveform with the above excitation structure, it is more efficient to change the ordering and extract $n=+5$ and $+6$ resonant modes before extracting the subdominant $n=+4$ overtone. 
This change of order effectively suppresses error propagation from the anomalously small $n=+4$ excitation.
Indeed, a comparison between the two extraction orders confirms that extracting $n=+5$ and $+6$ before $n=+4$ yields noticeably better accuracy: the relative errors of the resonant modes are reduced by about 20--50\%.

Of course, this discussion strictly applies to the ringdown induced by localized initial data, for which the mode amplitudes are determined solely by the excitation factors. 
In more general settings, one must also account for source- and initial-data-dependent effects. 
Nevertheless, given the magnitude of the hierarchy found here, it is reasonable to expect that the qualitative conclusion about the resonance-induced mode hierarchy would persist even when such effects are included.

\section{\label{sec:conclusion}Conclusion}
We identified and characterized how resonances near exceptional points manifest in Kerr ringdown waveforms.
We applied the iterative fitting method with mirror modes, supplemented by iterative subtraction of exact QNMs as a diagnostic check, to a ringdown gravitational waveform induced by a localized source term~\eqref{eq:deltafunctionsource1}.
A key advantage of employing this particular waveform is that its QNM amplitudes are directly given by the excitation factors, which have been recently computed and crossvalidated.
This controlled setup allowed us to investigate both mild and sharp resonances and to clarify their distinct imprints on the ringdown signal.

As a reference dataset (RD) of QNM frequencies and excitation factors, we used the one provided recently~\cite{Motohashi:2024fwt,motohashi_2024_12696858}.
Complementing the three previously established methods, we have provided an additional independent validation of the RD using the confluent Heun function (see Table~\ref{tab:efs}), thereby reinforcing its reliability and precision.
Employing these excitation factors, we demonstrated a direct and consistent agreement between the ringdown waveform generated by localized initial data and a superposition of QNM damped sinusoids.
This consistency establishes a solid basis for further studies of QNM excitations.

Building on this validated setup, we performed a detailed analysis of the resonance features in the ringdown.
In addition to the iterative fitting analysis, we also explored iterative subtraction of exact QNMs, which serves as a benchmark mimicking an ideal iterative fitting.
We focused on two representative cases: a mild resonance in $(l,m)=(2,2)$ at $a/M\simeq 0.9$, and a sharp resonance in $(l,m)=(3,1)$ at $a/M\simeq 0.9722$ occurring as part of a sequence of successive resonances.

For the mild resonance in $(l,m)=(2,2)$, we successfully extracted up to five QNM amplitudes.
We achieved the extraction of the fundamental mode and the first three overtones of the ordinary modes, and the fundamental mirror mode, all with an accuracy within 10\%.
The robustness of our extraction method is demonstrated by the presence of stable plateau regions, as shown in Fig.~\ref{fig:Stability}.
For higher overtones beyond the fourth, although stable plateau regions are still observed, the extracted amplitudes deviate by more than $10$\%, even though their overall magnitudes remain consistent with excitation factors. 
These deviations arise from the power-law tail and contamination due to error propagation from the extraction of lower overtones. 
As a result, our current setup is effectively limited to accurately extracting amplitudes up to around the third overtone.
To enable extraction of higher overtones, more reliable fitting algorithms and/or waveforms with weaker tail contributions will be required.

The comparison between residual waveforms after subtraction of extracted QNMs and exact QNMs provides further insights into the fitting process.
The damped sinusoids associated with the extracted QNMs are clearly observed in the residual waveform, except for the fifth overtone, whose decay times become similar at $a/M\simeq 0.9$.
While individual QNMs fail to match the residual damped sinusoid, the combined contribution of the fifth and sixth QNMs reproduces the observed oscillation.
This behavior results in relative errors of order unity in the extracted amplitudes of the fifth and sixth overtones.

For $(l,m)=(3,1)$, we found that the resonance causes a reversed hierarchy between overtone contributions in the ringdown signal.
Specifically, the iterative subtraction procedure clarified that the sharp resonance between $n=+5$ and $+6$ dominates, whereas the $n=+4$ mode provides only a minor contribution because of its exceptionally small excitation factor.
This suppression originates from the mild resonance between $n=+4$ and $+5$ prior to the sharp one between $n=+5$ and $+6$, leading to a rapid approach of the $n=+4$ excitation factor toward zero.
Thus, a lesson from our analysis is that resonance phenomena can not only \emph{amplify} QNM excitations but also \emph{reduce} them.

In summary, our analysis demonstrates characteristic contributions from resonance modes and elucidates their distinctive features in Kerr ringdown.
It would be interesting to apply improved fitting algorithms to mitigate the contamination and/or to reduce the degeneracy between overtones~\cite{Morisaki:2025gyu}.
It is also important to explore alternative waveform setups to achieve robust extraction of higher overtones.
We leave them for future work.

\begin{acknowledgments}
    We thank Naritaka Oshita for useful discussions.
    This work was supported by JST, the establishment of university fellowships towards the creation of science technology innovation, Grant Number JPMJFS2123 (K.K.) and Japan Society for the Promotion of Science (JSPS) Grant-in-Aid for Scientific Research (KAKENHI) Grant No.~JP22K03639 (H.M.).
    Numerical computation in this work was carried out at the Yukawa Institute Computer Facility.
\end{acknowledgments}

\section*{data availability}
The data that support the findings of this article are openly available~\cite{motohashi_2024_12696858,kubota_2026_18511201,kubota_2026_18676426}.

\appendix
\begin{widetext}
\section{\label{appendix:mirror}Rewriting mirror modes}
In the main text, we take into account the ordinary and mirror modes together.
The mirror modes are often rewritten in terms of the ordinary mode, but it seems there exists some confusion in the literature.
In this appendix, we provide a dictionary of the rewriting process.

For each $(l,m)$ mode, collecting the contribution from all the poles in the lower half complex plane, we obtain 
\begin{align}
    &  \sum_{n=-\infty}^{+\infty} C_{lmn} e^{-i\omega_{lmn}t} S_{lm}(\theta;a\omega_{lmn}) e^{im\phi}\nonumber \\
    &=  \sum_{n=+0}^{+\infty}  C_{lmn} e^{-i\omega_{lmn}t} S_{lm}(\theta;a\omega_{lmn})e^{im\phi}  +  \sum_{n=-\infty}^{-0}  C_{lmn} e^{-i\omega_{lmn}t}S_{lm} (\theta;a\omega_{lmn}) e^{im\phi} \nonumber\\
    &\eqqcolon \psi_\mathrm{ordinary} + \psi_\mathrm{mirror}.
\end{align}
As explained in the main text, we denote the ordinary mode by $+n$ and the mirror mode by $-n$. 
For $m>0$, they coincide with the positive-frequency and negative-frequency modes, respectively.
Since we have the relation~\eqref{eq:mirrormode} between the ordinary and mirror QNM frequency, $\omega_{lm-n}=-\omega^*_{l-mn}$, we can rewrite $\psi_\mathrm{retro}$ as the sum over the ordinary QNMs as 
\begin{align}
    \psi_\mathrm{mirror} &= \sum_{n=+0}^{+\infty}  C_{lm-n} e^{i\omega^*_{l-mn}t} S_{lm}(\theta;-a\omega^*_{l-mn}) e^{im\phi}
\end{align}
Next, we use the properties of the spin-weighted spheroidal harmonics function for spin-$s$ field [see Eq.~(48) in Ref.~\cite{Cook:2014cta}]:\footnote{It appears that the factor $(-1)^{l+s}$ is missing in Eq.~(3.6) in \cite{Press:1973zz}.}
\begin{align}
    {}_s S_{lm}(\theta;c) &= (-1)^{l+s}{}_s S_{l-m}(\pi-\theta;-c), \\
    {}_s S_{lm}(\theta;c^*) &= {}_s S_{lm}^*(\theta;c).
\end{align}
The present case for gravitational wave perturbation corresponds to $s=-2$, which we abbreviate throughout this paper. We thus obtain 
\begin{align}
    \psi_\mathrm{mirror} &= \sum_{n=+0}^{+\infty}  C_{lm-n} (-1)^l e^{i\omega^*_{l-mn}t} S^*_{l-m}(\pi-\theta;a\omega_{l-mn}) e^{im\phi}.
\end{align}
Finally, defining $C'_{lmn}\coloneqq C_{lm-n}(-1)^l$, we obtain 
\begin{align}
    \psi_\mathrm{ordinary} + \psi_\mathrm{mirror} &= \sum_{n=+0}^{+\infty} \left[ C_{lmn} e^{-i\omega_{lmn}t} S_{lm}(\theta;a\omega_{lmn})e^{im\phi}  +  C'_{lmn} e^{i\omega^*_{l-mn}t} \{ S_{l-m}(\pi-\theta;a\omega_{l-mn}) e^{i(-m)\phi} \}^*  \right].
    \label{eq:rewritemirrormode}
\end{align}
The final expression is consistent with expressions in the literature (see, e.g., Refs.~\cite{Berti:2005ys,Finch:2021iip,Isi:2021iql}).\footnote{It appears that $\omega_{lmn}$ in the second term in Eq.~(2.4) in Ref.~\cite{Dhani:2020nik} should be replaced by $-\omega_{l-mn}$, which is consistent with the present expression~\eqref{eq:rewritemirrormode}.}

\end{widetext}

\bibliographystyle{apsrev4-1mod}
\bibliography{references}

\end{document}